\documentclass[aps,prb,nopacs,floatfix,12pt]{revtex4}

\usepackage{graphicx}
\usepackage{amssymb}
\usepackage{amsmath}
\usepackage{epsfig}
\usepackage{bm}
\graphicspath{{/home/sgreben/figures/no2-kin/}}

\begin{document}
\bibliographystyle{jcp}

\title{Photodissociation of carbon dioxide in singlet valence
electronic states. 
II. Five state absorption spectrum and vibronic assignment}  

\author{Sergy\ Yu.\ Grebenshchikov\footnote{Sergy.Grebenshchikov@ch.tum.de}}
\affiliation{Department of Chemistry, Technical University of Munich,
  Lichtenbergstr. 4, 85747 Garching, Germany} 

\begin{abstract}
The absorption spectrum of CO$_2$  in the
wavelength range 120\,nm --- 160\,nm is analyzed 
by means of quantum mechanical
calculations performed using vibronically coupled PESs of five 
singlet valence electronic states and the coordinate dependent
transition dipole moment vectors. The thermally averaged spectrum, 
calculated for $T=190$\,K via
Boltzmann averaging of optical transitions from many initial rotational
states, accurtely reproduces the experimental 
spectral envelope, consisting of a low and a high energy band, 
the positions of the
absorption maxima, their FWHMs, peak intensities, 
and frequencies of diffuse structures in each band.  
Contributions of the vibronic interactions due to 
Renner-Teller coupling, conical intersections, and the Herzberg-Teller
effect are isolated and the calculated
bands are assigned in terms of adiabatic electronic states. Finally, 
diffuse
structures in the calculated bands are vibronically assigned using wave
functions of the underlying resonance states. It is demonstrated that
the main progressions in the high energy band correspond to
consecutive excitations of the
pseudorotational motion along the closed loop of the CI
seam, and progressions differ in the number of nodes along the radial mode
perpendicular to the closed seam. Irregularity of the diffuse peaks
in the low energy band is interpreted as a manifestation of the
carbene-type \lq cyclic' OCO minimum. 
\end{abstract}

\maketitle

\section{Introduction}\label{introduction}
This paper describes the results of an 
ab initio quantum mechanical study of the absorption spectrum
of carbon dioxide photoexcited with the
ultraviolet (UV) light between 120\,nm and 160\,nm and offers vibronic
assignments of the diffuse absorption bands. 
A brief account of this 
work has already been published.\cite{G12A} The potential
energy surfaces (PESs) used in these calculations are described in the
preceding paper (termed \lq paper I').\cite{G13A} 

Photodissociation of carbon dioxide is of considerable importance
for atmospheric and planetary chemistry. 
CO$_2$ is the second common trace gas in the Earth
atmosphere and one of the main products of the fossil fuel
burning. Its emissions continue to grow\cite{NOTE-CO2-0} 
despite considerable effort to mitigate climate 
change.\cite{UNFCCC,S13} UV photodissociation destroys
CO$_2$ with a unit quantum yield, and in the upper atmosphere, 
this destruction channel is dominant. Photolysis, shifted into the 
UV/visible range via preheating of CO$_2$, has been used in solar based
schemes of conversion of atmospheric CO$_2$ into fuel.\cite{TJ02,RVPG10} 
UV photochemistry of carbon dioxide controls chemical processes in the 
upper layers of the 
CO$_2$ based atmospheres of Mars and Venus.\cite{FBDD00} Accurate
predictions of the 
low temperature UV absorption cross sections are required to 
improve the existing photochemical models of their atmospheres, and are
also relevant for the models of the atmosphere of
Titan,\cite{VYC08} which contains CO$_2$ as a minor
constituent. Finally, photodissociation shares its activated
complex --- the
highly electronically and vibrationally excited CO$_2$ --- with another
reaction, in which $\rm C + O_2$ collisions produce carbon monoxide,  
the second most abundant molecule in interstellar clouds.\cite{GNC02}

Absorption spectrum of CO$_2$ in the UV range 
is studied experimentally since the beginning of the 20th 
century.\cite{L08,IWZ53,HERZBERG67,O71,RMSM71,CCB93,YESPIM96,SYSI07}
Figure \ref{spec1}(a) gives an overview between 220\,nm and 105\,nm.  
Carbon dioxide is transparent down to about
220\,nm (6.20\,eV). 
Below 220\,nm, the absorption starts with a very weak progression of
narrow lines.\cite{O71,CLP92,CLP05} At
112.1\,nm (11.08\,eV), a strong absorption band is observed, which is 
assigned\cite{RMSM71,CJL87} 
to the optically allowed transition to the Rydberg 
state $^1\Sigma_u^+$; this band is the first in a series of Rydberg
transtions at shorter wavelengths [not shown in Fig.\ \ref{spec1}(a)].
Between 160\,nm and 120\,nm (7.75\,eV --- 10.33\,eV), 
two relatively weak absorption bands are located, with the stronger
one centered at 133\,nm (9.30\,eV) and the weaker one at
148\,nm (8.40\,eV). 
Magnified view of this spectral interval is given in Fig.\
\ref{spec1}(b).  Each band exhibits irregular diffuse structures
superimposed on broad structureless background.
The vibronic origin of the diffuse structures 
and the dissociation mechanisms  in these two bands are subjects of
this paper. 

Previous analysis established key electronic states in the observed
spectrum.\cite{D63,HERZBERG67,RMSM71,EE79,KRW88,SFCRW91,SFCCRWB92} 
These are the
singlet valence states $2,3^1\!A'$ and  $1,2,3^1\!A''$, properties and 
topographies of the global PESs of which  are discussed in paper I. 
It is customary to give electronic assignment using 
the $D_{\infty h}$ labels, appropriate for linear CO$_2$, 
namely $1^1\Sigma_u^-$, $1^1\Pi_g$, and $1^1\Delta_u$, with the low energy band
at 148\,nm assigned to the $1^1\Delta_u$  state, and the
high energy 133\,nm band to the $1^1\Pi_g$ state.\cite{D63,HERZBERG67,RMSM71}  
However, the interstate vibronic couplings calculated in paper I
are strong rendering such \lq zeroth order' diabatic labels somewhat
problematic. Vibrational progressions are isolated using solely line spacings. 
For the stronger lines near 133\,nm, one
bending or two stretching progressions are offered  as
alternative tentative assignments.\cite{RMSM71,NOTE-CO21-1} 
The weaker band near 148\,nm is presumed to involve strong
bending excitations. This assignment is based on the
analysis of Dixon\cite{D63} who characterized the upper state in the 148\,nm
transition as bent (two-dimensional contour maps of the lowest two bent
excited states $2^1\!A'$ and $1^1\!A''$ are shown in Figs.\ 9 and 13 of
paper I). However, the intervals between diffuse peaks are 
strongly perturbed and rather irregular in both absorption bands: 
A unequivocal assignment is difficult 
without a convincing spectroscopic model or accurate
dynamics calculations.

Irregularities in the peak spacings and
intensity distributions are due to strong vibronic interactions, rooted
in the symmetry properties of the excited valence states. In the 
near linear Franck-Condon (FC) region, vibronic effects can be grouped
into three classes: \newline
\noindent{\bf (1)} The rotoelectronic Renner-Teller (RT) effect involves 
$A'$ and $A''$ components of the orbitally degenerate $1^1\Pi_g$ or
$1^1\Delta_u$ state. Figure \ref{copot1}(a) illustrates how the degeneracy
is lifted as CO$_2$ bends. In the shown cut, the upper two states stem
from $1^1\Pi_g$, the lower two stem from $1^1\Delta_u$, and the
transitions $1^1\Pi_g \leftarrow \tilde{X}^1\Sigma_g^+$ 
and $1^1\Delta_u \leftarrow \tilde{X}^1\Sigma_g^+$ can be nominally
classified as linear-linear and linear-bent, respectively.
In either case, the $A'/A''$ 
interaction  $\sim \Lambda\Omega/\sin^2\alpha_{\rm OCO}$
is proportional to the projections $\Lambda$ and $\Omega$ of the electronic 
($\hat{L}$) and  the total angular momentum 
($\hat{J} = \hat{N}_{\rm CO2} + \hat{L}$) 
on the molecular axis and, for $\Omega \ne 0$, diverges as
$\alpha_{\rm OCO}\rightarrow 180^\circ$.\cite{R34,HERZBERG67,P88,GGH93}  
If $\hat{J}$ is normal to the CO$_2$ axis and $\Omega = 0$, the RT
coupling is quenched. Interaction between $A'$ and $A''$
states in this case is mediated only by the
Coriolis coupling which allows transitions from 
$\Omega=0$ to $\Omega =1$ block in which RT coupling is operative. 

\noindent {\bf (2)} 
The symmetry allowed conical intersections (CIs) occur between
states $1^1\Delta_u$ and $1^1\Pi_g$; two avoided crossings 
in Fig.\ \ref{copot1}(b) near $2.2\,a_0$ and $2.9\,a_0$ illustrate
positions of the CIs. Calculations detect CIs for two $A'$ states,  
$2^1\!A'$   and $3^1\!A'$, and two $A''$ states, $1,2^1\!A''$  or
$2,3^1\!A''$.  In both symmetry blocks, the degeneracies are 
lifted linearly as the molecule bends or as one CO bond is extended. 
The two intersections, implied in Fig.\ \ref{copot1}(b), are 
not independent: They belong to a single \lq CI seam' 
in the $(R_1,R_2)$ plane of the CO bond distances  
in linear OCO. The CI seam forms a closed loop as
discussed in Sect.\,III\,C of paper I and in Ref.\ \onlinecite{GB12}. 
In fact, the closed CI seam is a
line of fivefold degeneracy which passes through the FC region.
Calculations discussed below demonstrate that the
closed seam substantially affects vibronic eigenstates located above
the CIs in the strong absorption band at 133\,nm. 

\noindent {\bf (3)} The Herzberg-Teller (HT) effect\cite{HERZBERG67} 
is seen as a 
strong coordinate dependence of the transition dipole moments (TDMs)
with $\tilde{X}$. The TDM vectors in the principal axes system are shown in
Fig.\ \ref{tdms} as functions the CO bond length and the OCO
angle.\cite{NOTE-CO24-1}  Although the
HT interaction neither requires nor involves a direct state intersection,
its effect upon the intensity distribution in the spectrum
is virtually indistinguishable from that of a CI in the FC region. 
Optical transitions from $\tilde{X}^1\Sigma_g^+$ to the states 
$1^1\Sigma_u^-$, $1^1\Pi_g$, and $1^1\Delta_u$ 
are forbidden in the $D_{\infty h}$ symmetry group. 
The TDM along the molecular axis, $\mu_y$, transforms according to the
irrep $\Gamma^{\mu_y}=\sigma^+_u$, 
while the components $\mu_x$ and $\mu_z$ are of
$\Gamma^{\mu_x}=\Gamma^{\mu_z}=\pi_u$ symmetry. 
Direct products\cite{KDC84,T88} 
$\Gamma^{\rm el}_i \times \Gamma^{\mu}\times \Gamma^{\rm el}_f$
of these irreps with the electronic symmetries $\Gamma^{\rm el}_i=\Sigma_g^+$ 
and $\Gamma^{\rm el}_f= \Sigma_u^-$, $\Pi_g$, or $\Delta_u$ 
are never totally symmetric and all TDMs at the FC point vanish. 
Transitions become vibronically allowed in distorted molecule, and the TDMs
directly depend on displacements in non-totally symmetric
coordinates. The observed absorption between 120\,nm and 160\,nm is
entirely due to this coordinate dependence. For
linear deviations from $D_{\infty h}$, active in the transitions are
the antisymmetric stretch $R_- = (R_1-R_2)/\sqrt{2}$
(irrep $\sigma_u^+$) and the bend $\alpha_{\rm OCO}$
 (irrep $\pi_u$). In linear CO$_2$, the components $\mu_{x,z}$ are 
antisymmetric with respect to $R_-$
as illustrated in Fig.\ \ref{tdms}(a,c). The component
$\mu_y$ grows linearly with
$\alpha_{\rm OCO}$, while $\mu_{x,z}$ are almost independent of the
OCO angle [Fig.\ \ref{tdms}(d,e,f)].
The TDM components corresponding to states $1^1\Sigma_u^-$ and $1^1\Delta_u$ 
are easily identified at linearity [Fig.\ \ref{tdms}(a,c)] 
by their small absolute values; the transition to 
$1^1\Pi_g$ dominates in $C_{\infty v}$. 
The coordinate dependence of the TDMs, 
violating both the Born-Oppenheimer and the FC
approximations,\cite{HERZBERG67,OS73} is commonly ascribed to the
\lq intensity borrowing' from an adjacent bright lender
state. The nearest bright state which might contribute to the
vibronically induced TDMs in Fig.\ \ref{tdms} 
is the Rydberg state $1^1\Sigma_u^+$ lying almost 2\,eV above $1^1\Pi_g$. 
As deviations from $D_{\infty h}$ grow, higher than linear expansion 
terms contribute, all TDM components become nonzero, and all 
states can be excited in the bent molecule  [Fig.\
\ref{tdms}(d,e,f)]. The $A'$ states 
are excited via the in-plane components $\mu_{y,z}$. The out-of-plane 
component $\mu_x$ mediates excitation of the $A''$ states. 
The symmetry of $\mu_{x,y,z}$ in Fig.\ \ref{tdms} implies that 
parallel (perpendicular) transitions excite vibrational states which
are symmetric (antisymmetric) with respect to interchange, $R_1
\leftrightarrow R_2$, of the two CO bonds. 

In this paper, the absorption spectrum of CO$_2$ is calculated 
using ab initio PESs of the first five excited electronic states, 
vibronic interstate couplings, and 
multistate quantum dynamics for the nuclei. The study was designed with
three goals in mind: To reproduce the experimental 
spectrum, to assign the calculated absorption bands, and 
to examine the relative contributions of the above non-adiabatic
interactions. These tasks are connected, because  
photoexcitation brings CO$_2$ into the
region of multiple degeneracies (Fig.\ \ref{copot1}), and 
all low lying states, populated via the HT effect, might be interlinked
through an intricate network of couplings created at the RT intersections
and the CIs. A realistic assignment should reflect the extent of this
mixing.  The outline of the paper is as follows: The
multistate quantum mechanical calculations of the absorption spectrum
and resonances in dissociation continua of the five excited electronic 
states are summarized in Sect.\
\ref{qmc}. The calculated spectrum is compared with experiment and
assigned in Sect.\ \ref{results}. Individual rotoelectronic
transitions are isolated in Sect.\ \ref{elass} allowing \lq coarse
grained' electronic assignment of the two calculated 
bands and a qualitative assessment of the role of RT coupling. 
Section  \ref{vibassh} assigns vibronic progressions in
the stronger high energy band and shows that nodal lines of resonance 
wave functions follow the topography of the upper adiabatic
PESs near the closed CI seam. Vibronic assignments of the low energy
band, much less definitive because of strong anharmonic and vibronic
couplings, are discussed in Sect.\ \ref{vibassl}. Weak absorption of
the state $1^1\Sigma_u^-$ is the subject of Sect.\ \ref{vibasssi}. 
Summary is given  in Sect.\ \ref{concl}.

\section{Quantum mechanical calculations}
\label{qmc}

\subsection{Molecular Hamiltonian}
\label{qmca}

The starting point for the calculations is the $5\times 5$ 
molecular Hamiltonian written in the basis of the electronic
states $\left|\Delta'\right.\rangle$, $\left|\Pi'\right.\rangle$, 
$\left|\Sigma''\right.\rangle$, $\left|\Pi''\right.\rangle$,  and
$\left|\Delta''\right.\rangle$, diabatized with respect to CIs at bent
geometries and in the FC region as described in Sect.\,IVB of paper I: 
\begin{equation}
\label{ham00}
{\bf H}_0 = {\bf T} + {\bf S}^T{\bf V}^a {\bf S}
= {\bf T} + {\bf V}^d \, ,
\end{equation}
with the nuclear kinetic energy matrix ${\bf T}$ and 
the potential matrix ${\bf V}^d$, related to the diagonal matrix 
of ab initio 
energies ${\bf V}^a$ via orthognal adiabatic-to-diabatic (ADT)
transformation ${\bf S}$ and consisting of a $2\times 2$ block of  $A'$
and a $3\times 3$ block of $A''$ states:
\begin{equation}
\label{diarep1}
{\bf V}^{d} = \left(\begin{array}{cc|ccc} 
V_{\Delta'} & V_{\Pi'\Delta'}  & & & \\
V_{\Pi'\Delta'} & V_{\Pi'}& & & \\ \hline
 & & V_{\Sigma''}& V_{\Sigma''\Pi''} & V_{\Sigma''\Delta''}\\
 & & V_{\Sigma''\Pi''} & V_{\Pi''} 
& V_{\Pi''\Delta''} \\
 & & V_{\Sigma''\Delta''} & V_{\Pi''\Delta''} & V_{\Delta''}\\
\end{array}\right) \, .
\end{equation}
All matrix elements are smooth functions of molecular coordinates; 
couplings of the accidentally degenerate states 
$\left|\Sigma''\right.\rangle$ and 
$\left|\Delta''\right.\rangle$ to $\left|\Pi''\right.\rangle$
are set equal, and the RT-like $\Sigma''/\Delta''$ interaction
is neglected. This representation will hereafter be referred to as
CI-diabatic; by construction, the
projection of $\hat{L}$ on the molecular figure
axis, $\hat{L}^2_z$,   
in the CI-diabatic basis has a definite value of $\Lambda = 0,
1,$ or 4. The kinetic energy operator $\hat{T}$  is specified in the
body fixed (BF) frame in Jacobi coordinates $(R,r,\gamma)$ 
in the form, originally suggested by
Petrongolo\cite{P88} and modified 
for numerical implementations by Goldfield et 
al.:\cite{GGH93}  
\begin{eqnarray}
\label{ham01}
\hat{T} & = & \left[\hat{T}_R + \hat{T}_r + 
(B_R+B_r)\left(\hat{T}_\gamma +
  \frac{\hat{J}_z^2+\hat{L}_z^2}{\sin^2\gamma}\right) + 
B_R\left(\hat{J}^2-2\hat{J}_z^2-\hat{L}_z^2\right)
\right] \nonumber \\
& & -\left[2(B_R+B_r)\frac{\hat{J}_z\hat{L}_z}{\sin^2\gamma}
-2B_R\hat{J}_z\hat{L}_z\right] + 
\left[B_R\hat{J}_+\hat{j}_-+B_R\hat{J}_-\hat{j}_+\right] \, .
\end{eqnarray}
Jacobi coordinates in this expression comprise the distance $R$ between one
oxygen atom and the center of mass of CO, the CO distance $r$, and the
angle $\gamma$ between the vectors ${\bf R}$ and ${\bf r}$. Operators 
$\hat{T}_R$, $\hat{T}_r$, and $\hat{T}_\gamma$ are the standard\cite{ZHANG99}
kinetic energies for these coordinates; the rotational constants 
$B_{R}= \left(2\mu_RR^2\right)^{-1}$ and 
$B_{r}= \left(2\mu_rr^2\right)^{-1}$ depend on the corresponding reduced masses
$\mu_{R,r}$. $\hat{J}$ denotes the total angular momentum, 
$\hat{J} = \hat{N}_{\rm CO2} + \hat{L}$  ($\hat{N}_{\rm CO2}$ is the
angular momentum of nuclei). $\hat{J}_z$ and $\hat{J}_\pm$ denote the BF $z$
projection and the ladder operators of $\hat{J}$, respectively.
 The BF $z$ axis is
assumed to run parallel to ${\bf R}$ 
(\lq R-embedding', Ref.\ \onlinecite{TS82}); finally, $\hat{j}_\pm =
\left(\hat{J}_z-\hat{L}_z\right)\cot\gamma \mp \partial/\partial\gamma$. 
In the basis of CI-diabatic states, the first bracket in Eq.\ (\ref{ham01}) 
applies to the vibronic dynamics in either
$A'$ or $A''$ symmetry block; the second bracket is the RT
interaction 
between $A'$ and $A''$ states sharing the same $\Lambda^2$; the thrid
bracket is the Coriolis interaction which includes both intra-symmetry
(through e.g. $\hat{J}_+\hat{J}_z$ terms) and inter-symmetry 
(through e.g. $-\hat{J}_+\hat{L}_z$ terms) couplings. 

Divergence  at $\gamma \rightarrow 0,\pi$
of the kinematic RT coupling matrix elements for states with $\Lambda^2 \ne 0$ 
is removed by switching from the definite symmetry states 
$\left|\Lambda'\right.\rangle$ and $\left|\Lambda''\right.\rangle$
to the RT-diabatic eigenstates $\left|\Lambda^\pm\right.\rangle$
of the $\hat{L}_z$ operator at linearity,\cite{GGH93}   
\begin{equation}
\label{rtdia}
\left|\Lambda^\pm\right.\rangle 
= \frac{1}{\sqrt{2}}\left(\left|\Lambda'\right.\rangle \pm 
i\left|\Lambda''\right.\rangle\right) \, ;
\end{equation}
Applied to Eq.\ (\ref{diarep1}), this transformation gives a
full CI/RT-diabatic representation, with 
all kinematic couplings replaced with potential matrix elements. 
The matrix ${\bf V}^d$ loses its block diagonal form and
comprises three  groups of elements: (1) Diagonal CI/RT-diabatic 
\lq potentials' $\frac{1}{2}(V_{\Lambda'}+V_{\Lambda''})$, (2) off-diagonal
vibronic couplings due to CIs involving states with $\Lambda_1\ne\Lambda_2$,
$\frac{1}{2}(V_{\Lambda_1'\Lambda_2'}+V_{\Lambda_1''\Lambda_2''})$, 
and (3) off-diagonal rotovibronic couplings
$\frac{1}{2}(V_{\Lambda'}-V_{\Lambda''})$ due to RT effect.
The state $\left|\Sigma''\right.\rangle$, vibronic
couplings for which are set equal to those of $\left|\Delta''\right.\rangle$
 and the RT effect is absent, 
becomes decoupled from other states in this approximation. 

The rotoelectronic basis in which the molecular Hamiltonian is set consists
of direct products 
$\left\{|\Theta^{J}_{M\Omega}\rangle\left|\Lambda^\pm\right.\rangle\right\}$ 
of the electronic CI/RT-diabatic functions and the rotational basis functions
$|\Theta^{J}_{M\Omega}\rangle$ expressed via Wigner D matrices; 
 $M$ denotes the space-fixed and $\Omega$ the body-fixed projection of
 $J$. In the actual implementation, this basis is further
 transformed by taking parity adapted 
linear combinations\cite{GGH93,ZHANG99} of basis functions with 
$\pm|l| = \pm(\Omega+\Lambda)$ or $\pm|l| = \pm(\Omega-\Lambda)$.
Further, the Jacobi angle is discretized and the
angular differential operators are set in the Gauss-associated
Legendre discrete variable representation (DVR). 
Finite basis representation is kept only in the quantum numbers
$\Omega$ and $\Lambda$. 

\subsection{The initial state and the absorption spectrum}
\label{qmct}

CO$_2$ in the rotational state $(J_i,\Omega_i)$ in
$\tilde{X}^1\Sigma_g^+$ is described by the wave function
\begin{equation}
\label{wf00}
\Psi_X({\bf q}|J_i,\Omega_i) =
\psi_0^{J_i,\Omega_i}({\bf q})|\Theta^{J_i}_{M\Omega_i}\rangle \, .
\end{equation} 
Here $\psi_0^{J_i,\Omega_i}({\bf q})$ is a given vibrational state 
and  ${\bf q}$ denotes 
three internal (e.g. Jacobi) coordinates. Since $\tilde{X}$ is a 
$\Sigma$ state, the initial total
angular momentum $J_i$ coincides with the nuclear angular momentum 
$N_{\rm CO2}$. Centrifugal sudden (CS) Coriolis-free 
approximation is used in Eq.\ (\ref{wf00}) treating CO$_2$ as a symmetric
top with conserved helicity $\Omega_i$. 
Only excitations from the vibrational ground state 
are considered, which for linear CO$_2$ 
implies $\Omega_i=0$, too. 

The state of CO$_2$ immediately after photoexcitation is first set in
the CI-diabatic representation. TDMs with $\tilde{X}$ 
are obtained from ab initio TDMs using the ADT matrix ${\bf S}$: 
\begin{equation}
\label{mudi}
{\bm \mu}^d = 
\left({\bm \mu}_{\Delta'}^d,{\bm \mu}_{\Pi'}^d,{\bm \mu}_{\Sigma''}^d,
{\bm \mu}_{\Pi''}^d,{\bm \mu}_{\Delta''}^d\right)^T = 
{\bf S}^T{\bm \mu}^a = {\bf S}^T
\left({\bm \mu}_{2A'}^a,{\bm \mu}_{3A'}^a,{\bm \mu}_{1A''}^a,
{\bm \mu}_{2A''}^a,{\bm \mu}_{3A''}^a\right)^T
\, ,
\end{equation}
The resulting TDM \lq vector' has five electronic components, 
each featuring up to two non-vanishing spatial projections in the 
principal axes frame.  
Possible rotational transitions in the optical excitation are $J_i
\rightarrow J_f = J_i,J_i \pm 1$ and $\Omega_i=0 \rightarrow \Omega_f =
0,1$ (parallel case) or 
$\Omega_i=0 \rightarrow \Omega_f = 1$ (perpendicular case). 
Parallel and perpendicular transitions are calculated
separately. The electronic component of the 
initial excitation in the CI-diabatic
state $\Lambda$ with the angular momentum $J_f$ and  
the BF projection $\Omega_f$ reads as
\begin{equation}
\label{iniB}
\Phi({\bf q}|J_f,\Omega_f)_\Lambda = \langle 1 \delta,\,J_i
0|J_f\Omega_f\rangle\,
(\mu^d_\Lambda)_\delta\, \psi_0^{J_i,0}({\bf q})\,
|\Theta^{J_f}_{(m+M)\Omega_f}\rangle
\, ,
\end{equation}
Here $\langle \cdot,\cdot\,|\,\cdot\rangle$ is the Clebsh-Gordon coefficient, 
$\delta \equiv \Omega_f-\Omega_i = 0$ or 1, 
and $(\mu_\Lambda)_\delta$ are the spherical projections of the TDM vectors 
${\bm \mu}_\Lambda$: $(\mu)_{\delta=0} = \mu_z$ and $(\mu)_{\delta=+1} 
= -\left(\mu_y-i\mu_x\right)/\sqrt{2}$. 
This expression is valid for both parallel and perpendicular
transitions. The only difference between the two is that the 
initial state in a parallel
transition is real, while in a perpendicular transition it is imaginary.
For each type of transition, the total initial wavefunction 
$\Phi({\bf q}|J_f)$ is a
linear combination of all $\Phi({\bf q}|J_f,\Omega_f)_\Lambda$ with the allowed
values of $\Lambda$ and 
$(J_f,\Omega_f)$.\cite{BK04} Initial state with each
$J_f$ is propagated separately. The corresponding
cross section, averaged over the initial space fixed projections $M$, 
is proportional to the expectation value of the Green's function
of the time-independent Schr\"odinger equation,\cite{MT97} 
\begin{equation}
\label{cross00}
\sigma(E_{\rm ph}|J_f \leftarrow J_i) = 
\frac{\pi E_{\rm ph}}{3c\epsilon_0\hbar}
\langle \Phi({\bf q}|J_f)|{\rm Im} G^+(E_{\rm ph})| \Phi({\bf
  q}|J_f)\rangle \, , 
\end{equation}
where the photon energy $E_{\rm ph}$ is measured from the 
zero point energy level $E_0(J_i,\Omega_i=0)$ in $\tilde{X}$. 
The total cross section is the
sum over all allowed $J_f$:  
\begin{equation}
\label{cross01}
\sigma(E_{\rm  ph}|J_i) = \overline{\sum}_{J_f}\sigma(E_{\rm
  ph}|J_f \leftarrow J_i)  \, .
\end{equation}
The calculations including all $\Omega_f$
blocks for each $J_f$ and thus fully including
Coriolis coupling between adjacent helicity blocks were
performed for $J_i \le 2$. 

In order to compare the calculation with experiment at $T = 190\,K$, 
a thermal absorption spectrum has been constructred
by averaging $\sigma(E_{\rm ph}|J_i)$
over the normalized 
Boltzmann population  $P_{J_i}(T)$ of the initial states $J_i$, 
\begin{equation}
\label{thersi}
\sigma(E_{\rm ph},T) = \sum_{J_i}P_{J_i}(T)\sigma(E_{\rm
  ph}|J_i) \, . 
\end{equation}
Because of  large $J_i \le 45$ involved in these calculations, CS
approximation was applied,  the
Coriolis coupling neglected, and each initial state
$\Phi({\bf q}|J_f,\Omega_f)$ processed separately. 
For $J_i \le 2$, the spectra obtained with and
without CS approximation are in excellent agreement.

In practice, calculations of the absorption spectrum via Eq.\
(\ref{cross00}) 
are performed in several steps. The CI-diabatic initial state 
$\Phi({\bf q}|J_f)$ is transformed, first to the CI/RT-diabatic and then to
the parity adapted representation.
The last transformation is dropped in the CS approximation. 
Next, the outgoing boundary conditions are 
approximately imposed in the asymptotic region by augmenting 
$\hat{\bf H}_0$ with a complex absorbing potential $-iW$ (Re\,$W >
0$):\cite{JA85}
\begin{equation}
\label{ham02}
\hat{\bf H} = \hat{\bf H}_0 - i\,W \,{\bf 1} \, .
\end{equation}
Here ${\bf 1}$ denotes the $5\times 5$ unit matrix, and one
coordinate function $W$ is used in all electronic channels. 
In the last step, the vector 
${\rm Im}\,G^+(E)\,\Phi({\bf q}|J_f)$ is approximated by the
expansion
\begin{equation}
\label{mcheb01}
{\rm Im}\,G^+(E)\,\Phi({\bf q}|J_f)\simeq \sum_{n = 0}^{N_{\rm iter}}
[{\rm Re}\,b_n(E)]\,\Phi_n \, ,
\end{equation}
in which 
$b_n(E)$ are the usual energy-dependent coefficients,\cite{TK84} 
while the vectors $\Phi_n$ are found from a modified Chebyshev 
recursion relation due to Mandelshtam and Taylor.\cite{MT95}
Chebyshev autocorrelation coefficients $c_n =
\langle\Phi_0|\Phi_n\rangle$ with the initial
state $\Phi_0 \equiv \Phi({\bf q}|J_f)$ are calculated, and 
the absorption spectrum is reconstructed from $\{c_n\}$ 
using the Eq.\ (\ref{cross00}).  
The number of expansion terms, $N_{\rm iter}$, is one
of the convergence parameters of the calculation. 

In order to assign diffuse peaks in the spectrum, 
filter diagonalization\cite{WN95,MT97}
has been performed. Eq.\ 
(\ref{mcheb01}) is used to generate a set of basis vectors $\{\psi_k\}$, 
\begin{equation}
\label{fd01}
\psi_k = {\rm Im}\,G^+(E_k)\,\Phi_0 \, , 
\end{equation}
localized around nodes $\{E_k\}$ of an equidistant energy grid
covering a given window $[E_{\rm min},E_{\rm max}]$
(500\,cm$^{-1}$ wide windows were used in most calculations). 
The Hamiltonian matrix $\langle \psi_k|H|\psi_{k'}\rangle$ is set in
the orthogonalized basis and diagonalized. 
The resulting complex eigenstates are resonance states of the Hamiltonian
(\ref{ham02}) with energy $E_n$ and width $\Gamma_n$. 
The eigenfunctions $\Psi_n$ are five component vectors known in any
representation used in the calculations. For example, the CI-diabatic
eigenstate, 
${\bm \Psi}_n
=\left(\Psi^{\Delta'}_n,\Psi^{\Pi'}_n,\Psi^{\Sigma''}_n,\Psi^{\Pi''}_n,\Psi^{\Delta''}_n\right)^T$,
can be rewritten in the CI-adiabatic representation using the
ADT matrix via ${\bf S}^T{\bm \Psi}_n$. 

\subsection{Numerical details}
\label{qmcb}

Quantum mechanical calculations have been performed with several
sets of the diabatized PESs constructed using slightly different
fitting procedures for ADT parameters as described in Sect.\ IVB of 
paper I. All diabatic sets give indistinguishable absorption spectra. 
Hamiltonian is set in DVR\cite{LC00} with the grid comprising 
150 potential-optimized\cite{EC92} points in 
$R \in [1.5,7.0]\,a_0$, 100 potential-optimized points in 
$r \in [1.5,6.5]\,a_0$, and $N_\gamma = 100$ 
Gauss-associated Legendre quadrature points in angle. 
The grid is contracted by retaining only the points
with potential energy below $V_{\rm cut} = 12.5$ eV above
equilibrium in $\tilde{X}$. $N_{\rm iter} = 15\,000$ Chebyshev
iterations are sufficient to converge the absorption spectrum. 
The complex absorbing potential $-i\,W$ is set via a coordinate
dependent damping function $\gamma$,\cite{MT95} which is zero in the
inner parts of the five adiabatic potentials and grows quadratically
as a function of CO$\cdots$O distance 
in the asymptotic regions. The quadratic polynomial
$\gamma(Q)$ ($Q = R$ or $r$) is described in Refs. 
\onlinecite{MT97} and \onlinecite{KASGSCH00}. 
It becomes non-zero 1.1\,$a_0$ before the grid edge and reaches the
strength of 0.15 at the edge. 

Temperature dependent absorption cross section is
based on spectra calculated for all initial $J_i$ between 0 and
45, the range 
sufficient for temperatures up to at least $T \sim 250$\,K.  

Filter diagonalization is performed in several overlapping energy
windows covering the photon energy interval from
60\,000 cm$^{-1}$ to 80\,000 cm$^{-1}$.
Calculated resonances are broad, with the average width of
$\overline{\Gamma}_n \ge 50$ cm$^{-1}$, and require 
25\,000 Chebyshev interations for convergence. As discussed in Sect.\
\ref{vibassh}, the calculated resonances allow a qualitatively
accurate reconstruction of the absorption spectrum. 
All quantum mechanical calculations in this work are performed using the
program package \lq PolyWave'.\cite{NOTE-CO21-4}

\section{Results}
\label{results}

The absorption spectrum, thermally averaged at $T=200$\,K,  is
compared with the experimental cross section in
Fig.\ \ref{spec1}(b). The calculated spectrum has a characteristic 
two-band shape, and the band centers at
8.43\,eV (147.2\,nm) and 9.28\,eV (133.6\,nm)
are accurate within 0.05\,eV. 
The maximum cross sections in the two bands, 
$5.0 \times 10^{-19}$\,cm$^{2}$ and $9.5 \times 10^{-19}$\,cm$^{2}$, 
are about 15\% below the experimental values, but the observed 1:2 
intensity ratio between the weaker and the stronger band
is accurately reproduced. In both spectra, 
structures in the weaker 148\,nm band are substantailly more
diffuse and less regular 
than in the better resolved stronger  133\,nm band. 
Large discrepancies are found in the spectral widths (FWHMs). 
For the low and the high energy band, 
FWHMs are $7500$\,cm$^{-1}$ and $6100$\,cm$^{-1}$, i.e. 
off by 6\% and 18\% as
compared to the respective values of 7000\,cm$^{-1}$ and 
7400\,cm$^{-1}$ obtained from deconvolution of the experimental
spectrum of Ref.\ \onlinecite{CCB93}. 
The exaggerated gap between the two calculated bands
is one of the reasons for this deviation. The gap depth, only weakly 
dependent on diabatization details,\cite{G12A,G13A} is
sensitive to the TDM derivatives with respect to molecular coordinates
at the FC point. In several test calculations, these derivatives --- the
manifestations of the HT effect --- were artificially modified, and
the angular derivative 
$\partial\,\mu_y/\partial\,\alpha_{\rm OCO}$ had the largest impact
on the band intensities and the gap size.

The impact of thermal averaging can be assessed by comparing the $200$\,K 
spectrum in Fig.\ \ref{spec1}(b) with the \lq 0\,K'
spectrum, $\sigma(E_{\rm  ph}|J_i = 0)$,  calculated for 
a single initial rotational state $J_i = 0$ and shown  in Fig.\
\ref{spec2}(a). The differences in
positions and maximum intensties are marginal: The 200\,K spectrum is red 
shifted by $\sim 100$\,cm$^{-1}$ 
and its intensity is about 12\% higher. Boltzmann
averaging primarly affects the extent of spectral fluctutations
which in the 200\,K spectrum is substantially attenuated.  
The reason is the inhomogeneous broadening
in the spectrum \lq assembled' from an incoherent
average over many spectra for different rotational states
$J_i$. Narrowly spaced and erratically fluctuating lines 
in the low energy band are more amenable to 
thermal \lq washing out' than the well
resolved peaks in the high energy band, and the temperature effect is 
most conspicuous below 9\,eV.

\subsection{Electronic assignments}
\label{elass}

The currently prevailing assignment, attributing the 148\,nm band to
$1^1\Delta_u$ state and the 133\,nm band 
to $1^1\Pi_g$ state,\cite{HERZBERG67,RMSM71,KRW88} is somewhat
ambigous because these states cross in the FC region. Moreover, this
assignment clearly refers to linear geometries because 
CI-diabatic picture is explicitly selected as a reference. 
Quantum mechanical approach offers direct means to
clarify the electronic origin of each band. For example, the
assignments in Ref.\ \onlinecite{G12A} were determined  using
population patterns of the electronic components of resonance 
wave functions ${\bm \Psi}_n$. The results indicate that it is
the CI-adiabatic representation which is adequate for the assignment of
resonances,  
implying a strong vibronic coupling scenario:\cite{GSCHQZ06A}
The $2^1\!A'$ and $1^1\!A''$ adiabatic components are strongly populated 
below the CIs, while the components $3^1\!A'$ and $3^1\!A''$ become
dominant above the CIs. However, the population analysis 
works best if the electronic states involved are all
bound or all unbound. If, as in the present case, the molecule is
excited into the energy range in which the linear states $3^1\!A'/3^1\!A''$
are effectively bound, while the bent states 
$2^1\!A'/1^1\!A''$ are dissociative, the
results might become sensitive to the
precise definition of the \lq inner region' used to
evaluate electronic populations. 

The CI-adiabatic assignments of Ref.\ \onlinecite{G12A} can be corroborated
using  short-time dynamics --- a natural approach to establish
the electronic labels which refer to the 
largest energy scale (i.e. the smallest time
scale) in the molecule. To this end, two auxiliary diabatic TDM vectors 
are constructed using the ADT of Eq.\ (\ref{mudi}), 
one exclusively from the TDMs of the bent adiabatic states 
$2^1\!A'/1^1\!A''$ (termed ${\bm  \mu}^d_{21}$) 
and the other exclusively from the TDMs of the linear adiabatic
states $3^1\!A'/3^1\!A''$ (termed ${\bm  \mu}^d_{33}$):
\begin{eqnarray}
\label{mudi2}
{\bm \mu}_{21}^d & = & {\bf S}^T
\left({\bm \mu}_{2A'}^a,\,\,\,\,0\,\,\,\,, {\bm \mu}_{1A''}^a,
0,\,\,\,\,0\,\,\,\,\right)^T\\
{\bm \mu}_{33}^d & = & {\bf S}^T  
\left(\,\,\,\,0\,\,\,\,, {\bm \mu}_{3A'}^a,\,\,\,\,0\,\,\,\,,
0,{\bm \mu}_{3A''}^a\right)^T
\, .
\end{eqnarray}
 For each auxiliary diabatic TDM, the initial excitation, constructed
using Eq.\ (\ref{iniB}) for $J_i = 0$, is propagated under
the full Hamiltonian for 500 Chebyshev iterations corresponding to a
time interval of $\sim 0.1$\,fs.  
The resulting low resolution spectra are shown in the upper part of Fig.\
\ref{spec2}(b). They provide a pictorial decomposition of the full spectrum
$\sigma(E_{\rm  ph}|J_i = 0)$ 
into the components due to excitations via ${\bm  \mu}^d_{21}$
and ${\bm  \mu}^d_{33}$. Although not quantitative, the result
is clearly consistent with the previous assignment of the 148\,nm
band to the bent pair $2^1\!A'/1^1\!A''$ and the 133\,nm
band to the linear pair $3^1\!A'/3^1\!A''$.

The CI-adiabatic assignment is also motivated by the shape of the
initial excitation along the bending angle,
sketched in Fig.\ \ref{copot1}(a) together with the cuts through the 
calculated PESs. The dashed red line is the ground vibrational state in
$\tilde{X}$. Its maximum is shifted by 4$^\circ$ away from
linearity due to the zero-point bending vibration plus the weighting
imposed by the volume element. 
Thus, even before the TDM with the excited states is applied to
$\psi_0^{1,0}({\bf q})$,  the FC geometry corresponds to 
$\alpha_{\rm OCO} = 176^\circ$, and the 
gap between the adiabatic linear and bent excited states is $\sim
0.6$\,eV.  The
solid red line in Fig.\ \ref{copot1}(a) exemplifies the influence of the
TDM. Shown is the initial excitation created in the state $3^1\!A'$ via 
the TDM $\mu_y$  (the initial excitation in $2^1\!A'$ looks very
similar).  $\mu_y$ strongly depends on angle because of the HT
effect, as shown in Fig.\ \ref{tdms}(e). As a result, the non-vertical
transition bends the molecule even stronger, shifting the 
probability maximum to $\alpha_{\rm OCO} =  173^\circ$. 
The linear/bent gap at this angle becomes 0.86\,eV and matches the
energy separation between the maxima of the two absorption 
bands (0.85\,eV). In
fact, the adiabatic gap is a primary control handle of the
positions of the band maxima: If the original 
potentials are modified to reduce the gap size step by step, the
bands gradually approach each other and eventually merge.

While vibronic couplings and the HT effect directly 
shape the UV absorption of CO$_2$, the role of the rotoelectronic 
RT coupling is more subtle. 
The lower part of Fig.\ \ref{spec2}(b) shows 
the parallel, $\sigma_\parallel$,  and perpendicular, $\sigma_\perp$, 
components of the full spectrum $\sigma(E_{\rm  ph}|J_i = 0)$.   
In the parallel transition, both $\Omega_f=0$ and 
$\Omega_f=1$ states are initially populated in the $A'$ symmetry
states, but 86\% of the population is in
the $\Omega_f=0$ block reached via the strongly angular dependent TDM
$\mu_y$. RT coupling is zero for $\Omega_f=0$; it is active only
in the  almost empty 
$\Omega_f=1$ block, reached via $\mu_z$, and is promoted by the 
Coriolis-induced population transfer from $\Omega_f = 0$ to
$\Omega_f = 1$. If the molecule dissociates exclusively out of $\Omega_f =
1$ states, Coriolis transitions 
control the dissociation rate.  Dissociation of 
HCO in the $\tilde{A}^2\!A''$ state is a
well known example.\cite{GGH93,WSCHM00} In contrast, dissociation of
CO$_2$ is fast even for $\Omega_f = 0$ and the Coriolis interaction is
inefficient on the characteristic time scale of
$\tau_{\rm diss} \sim \hbar/\overline{\Gamma}_n \sim 100$\,fs. 
This is also the reason why  CS approximation has little effect on the
parallel spectrum $\sigma_\parallel$.

The perpendicular transition terminates in $A''$ states and is realized
via TDM $\mu_x$. The spectrum $\sigma_\perp$ in Fig.\
\ref{spec2}(b) consists of two widely separated almost equally intense
bands featuring apparently regular
progressions of diffuse peaks. The reason for the distinct
regularity of the 148\,nm band in $\sigma_\perp$ will
be discussed in Sect.\ \ref{vibassl}. The rotational state
$(J_f=1,\Omega_f=1)$ is exclusively populated, and the 
RT coupling is operative in the perpendicular
transition. However, $\sigma_\perp$ is
substantially smaller than $\sigma_\parallel$ for all but the highest
photon energies: The TDM $\mu_x$, which is similar to $\mu_z$,  depends weakly 
on $\alpha_{\rm OCO}$ (see Fig.\
\ref{tdms}), and it is excitations from bent geometries which dominate the
spectrum. As a result, the RT coupling, active in $\sigma_\perp$,
has little influence on the sum $\sigma_\perp +
\sigma_\parallel$.\cite{NOTE-CO24-2} 

The state $^1\Sigma_u^-$, decoupled from other states in this
calculation, is excited in a perpendicular transition via $\mu_x$. 
Its absorption spectrum, discussed in Sect.\ \ref{vibasssi}, 
consists of a series of sharp narrow lines, with the peak intensity
reached in the gap between the two main bands. The TDM
with $\tilde{X}$ is extremely small in the FC region, and 
the absorption cross section does not exceed
$5\times10^{-20}$\,cm$^2$. As a result, the contribution of 
the $^1\Sigma_u^-$ state is hardly discernible in the full spectrum.
Test calculations register minor intensity gain 
if weak vibronic coupling to other states is included. 

\subsection{Vibronic assignments of the high energy band}
\label{vibassh}

The intense lines in the experimental 133\,nm band [Fig.\ \ref{spec1}(b)] are
spaced, on average, by $\sim 630$\,cm$^{-1}$. This major
progression is shown by brown sticks in Fig.\ \ref{spec2}(a). 
Another progression can be recognized as minor peaks or shoulders
above 72\,000\,cm$^{-1}$. 
The minor peaks, also shown as sticks in Fig.\ \ref{spec2}(a), are 
red shifted by $\sim 250$\,cm$^{-1}$ with respect to the major
ones. Above 77\,000\,cm$^{-1}$, the diffuse bands consist of two
clearly separated approximately equally intense peaks.
Despite its simple appearance, even the major
progression resists straightforward assignment because 
the spacings between the adjacent peaks are not
constant. Instead, they obey an intricate rise-and-fall 
pattern in which intervals of
frequency growth (for example, 73\,000\,cm$^{-1}$---75\,000\,cm$^{-1}$
or 76\,000\,cm$^{-1}$---78\,000\,cm$^{-1}$)
are interrupted by dips (for example, near
73\,000\,cm$^{-1}$, 75\,300\,cm$^{-1}$, or
79\,000\,cm$^{-1}$). Rablais et al. suggested that the
progression is due to bending excitations in the near-linear molecule,
with the perturbations originating from the Coriolis interaction
between the vibrational and rotational angular momenta.\cite{RMSM71}
The dynamics calculations suggest that the major progression is built
on synchronized excitations of the two CO bonds.  

In the calculated spectrum [Fig.\ \ref{spec1}(b) and \ref{spec2}(a)], 
the major and minor progressions are qualitatively 
similar to the experimental ones, but numerically
different. For example,  the spacings between major peaks follow
the same rise-and-fall pattern, but their average is merely $\sim
580$\,cm$^{-1}$.  Next, the minor progression, also
seen in the calculations,  is blue-shifted by $\sim
240$\,cm$^{-1}$ against the major one. Around
76\,000\,cm$^{-1}$, the intensity of the minor peaks grows substantially
and the calculated spectrum shows several strong lines giving it 
a congested appearance.  Above 77\,000\,cm$^{-1}$, most calculated
peaks are split into two components in agreement with
experiment. Thus, the achieved 
accuracy is not spectroscopic, but the calculated spectrum
represents a reasonable starting point for the analysis of 
molecular motions behind the diffuse bands. The assignment, presented below, 
is based on the properties of metastable resonance states 
calculated specifically for the  parallel component $\sigma_\parallel$ 
of the full spectrum. The principal progressions are similar in 
 $\sigma(E_{\rm  ph}|J_i = 0)$ and 
 $\sigma_\parallel(E_{\rm  ph}|J_i = 0)$ [cf. Fig.\ \ref{spec2}(a) and
 (b)]. Differences visible above 78\,000\,cm$^{-1}$ are 
addressed at the end of this section.

Positions, $E_n$, of resonance states excited in a parallel transition 
$(J_i = 0,\Omega_i = 0) \rightarrow (J_f = 1,\Omega_f = 0)$,
calculated using filter diagonalization in the
CS approximation, are illustrated with sticks in Fig.\
\ref{spec2}(c). Each resonance is further 
characterized by its 
width $\Gamma_n$ and the intensity  $I_n = \left|\langle {\bm
\Phi}(J_i,\Omega_i)|{\bm   \Psi}_n\rangle\right|^2$. 
The resonance absorption spectrum, 
\begin{equation}
\label{sigmares}
\sigma_{\rm res}(E_{\rm ph}|J_f,\Omega_f) = 
\sum_{n}I_n\frac{\Gamma_n^2/4}{(E_{\rm ph}-E_n)^2+\Gamma_n^2/4} \, ,
\end{equation}
is compared with $\sigma_\parallel$ in Fig.\ \ref{spec2}(c). 
Resonances are expected to accurately describe 
absorption spectra consisting of narrow isolated lines.\cite{GSCHH03} 
However, despite a substantial fast direct contribution, the incoherent sum
of  Lorentzians in $\sigma_{\rm res}$
is in qualitative and --- above 73\,000\,cm$^{-1}$ --- even in
quantitative agreement with $\sigma_\parallel$, both in terms of 
positions of the diffuse lines and the intensity of the background.  

Nodal patterns of resonance wave functions provide
vibronic labels for the diffuse peaks. All resonances with
$\Gamma_n \le 150$\,cm$^{-1}$ were screened, but the discussion will be
limited to states with the largest
intensities. Their positions are shown with sticks in
Fig.\ \ref{spec2}(d), and their widths range from 30\,cm$^{-1}$ to 
80\,cm$^{-1}$.  Relevant for the assignment is the $3A'$ 
electronic component of the ${\bm \Psi}_n$ 
vector in the CI-adiabatic representation. 
Indeed, the 133\,nm band is due to the linear adiabatic states 
$3^1\!A'/3^1\!A''$.  Moreover, the considered
parallel transition populates only $A'$ states, with the population
transfer into the $A''$ block  suppressed 
for $\Omega_f = 0$ without Coriolis coupling.
Direct dissociation in the state $3^1\!A'$ is hindered
by a  $\sim 1$\,eV high barrier [see Fig.\ 1 of paper I], and
the $3A'$ component of ${\bm \Psi}_n$ describes a 
vibrational level trapped in the upper adiabatic sheet of the cone
and  non-adiabatically coupled to the dissociation continuum of 
the lower $2A'$ sheet (cf. Ref.\ \onlinecite{S63}). 

States in the major progression, which can be followed from the
origin near 72\,700\,cm$^{-1}$ up to at least 78\,000\,cm$^{-1}$,
are illustrated in the left column of Fig.\ \ref{wf01}. 
The 3D wave functions are viewed in the $(R_1,R_2)$ plane against the 
contour map of the $3A'$ potential containing the closed CI seam. 
Their assignment 
in terms of three quantum numbers $(v_\phi,v_r,v_b)$ is indicated
above each frame and the respective combs are shown in Fig.\
\ref{spec2}(d). These quantum numbers reflect 
three features shared by the states in the  main progression. 
First, they have no nodes along the bending direction except at
$\alpha_{\rm OCO} = 180^\circ$; most probable angles lie 
between 173$^\circ$ and 177$^\circ$ and the states are essentially \lq
flat', i.e. localized in the $(R_1,R_2)$ plane.  
Such resonances are labelled by $v_b = 0$. 
Second, these states have no
nodes along the direction normal to the CI seam, and their \lq radial'
quantum number is zero, $v_r=0$. Third, these states decribe
consecutive excitations around the CI loop. In each successive
state, two new nodes are \lq stringed' on the CI seam \lq wire', and the
quantum number for this mode, $v_\phi$, counts nodes
along one side of the seam. All states in Fig.\ \ref{wf01} are
symmetric with respect to interchange of $R_1$ and $R_2$, as expected
for parallel transitions; antisymmetric states, with a node along the
$R_1=R_2$ line, 
remain \lq invisible'. The potential along the seam loop is rather
smooth, and the average frequency in the $v_\phi$ progression is 
$\omega_\phi \sim 580$\,cm$^{-1}$. However, energies in 
the \lq angular' $v_\phi$ progression are not equidistant
(see below). The quantum numbers $v_r$ and $v_b$ correspond to the
tuning and the coupling modes of the CI, respectively.  The  
adiabatic potentials along these directions are cusped, and the 
vibrational frequencies $\omega_r$ and $\omega_b$ are expected to be
much larger than $\omega_\phi$. 

A simple
two-dimensional (2D) reference model, which illustrates the impact of the
closed CI seam on the eigenstates and the energy spectrum of the full
Hamiltonian, consists of a single adiabatic $3^1\!A'$ PES in the 
 $(R_1,R_2)$ plane with fixed $\alpha_{\rm OCO} = 179^\circ$. The
2D spectrum has been calculated using filter diagonalization, and 
examples of the 2D eigenstates are shown in the middle column of Fig.\
\ref{wf01}. They can be assigned two
 quantum numbers $(v_\phi,v_r)_{\rm 2D}$  
 measuring excitations along the angular seam mode $\phi$ and the radial
stretching  mode $r$. As in the full Hamiltonian, the states have
either a maximum or a node along the line $R_1 = R_2$. Symmetric  
wave functions in the progression $(v_\phi,0)_{\rm 2D}$ in Fig.\ \ref{wf01}
can be directly compared with  states in the major
progression $(v_\phi,0,0)$ in the left column. The 2D states are 
tightly localized around the CI loop, 
emphasizing that the quantum number $v_\phi$ describes
hindered rotation in the $(R_1,R_2)$ plane. Similarities in the
wave functions are strong enough to make the 2D model useful in disentangling
complicated nodal patterns in the full calculations. A good case in
point is the progression $(v_\phi,1)_{\rm 2D}$ built on one quantum of
radial excitation and shown in the lower panels of Fig. \ref{wf02}.  
The frequency of the radial mode, $\omega_r =
2150$\,cm$^{-1}$, is almost four times the
hindered rotation frequency. Resonance anharmonic couplings mix the states
$(v_\phi,0)_{\rm 2D}$ and $(v_\phi-4,1)_{\rm 2D}$, but their wave
functions remain clearly recognizable in 2D (compare states marked
with stars in Fig.\ \ref{wf01} and \ref{wf02}). 
The progression of states with $v_r = 1$ is found in the full calculation,
too (Fig.\ \ref{wf02}, upper panels). The fundamental 
radial frequency, $\omega_r = 1560$\,cm$^{-1}$, is
lower than in 2D. Nevertheless, 
the anharmonicity is strong and many states $(v_\phi,0,0)$ and
$(v_\phi-4,1,0)$ are mixed making their assignment difficult.  
In fact, the states in the upper panels of
Fig.\ \ref{wf02} were identified using their 2D analogues. Moreover,
most states $(v_\phi,1,0)$ are 
discenrible merely as satellites to the major peaks $(v_\phi,0,0)$; 
their intensity increases only above 78\,000\,cm$^{-1}$ 
[see Fig.\ \ref{spec2}(d)]. As a result, 
many members of this progression are missing, but the 
gaps can be filled in by the 2D model (cf. Fig.\
\ref{wf02}). 

The third progression clearly visible in the calculated spectrum 
 consists of resonances with a node at a bending angle
$\alpha_{\rm OCO} < 180^\circ$.  Such states are assigned
$(v_\phi,v_r,1)$. The state $(0,0,1)$ lies at 
75\,300\,cm$^{-1}$, so that the bending frequency is 
$\omega_b = 2290$\,cm$^{-1}$. 
The intensity in this progression quickly grows with
energy. Of the three strong bands, \lq congesting' the spectrum
between 75\,000\,cm$^{-1}$ and
76\,000\,cm$^{-1}$, two are due to states $(2,0,1)$ and $(3,0,1)$; only
one state, $(6,0,0)$, has no bending excitation. 
Above $76\,000$\,cm$^{-1}$ the states $(v_\phi,0,1)$ have
the largest intensities. Their
wave functions, shown in the right column of 
Fig.\ \ref{wf01}, are similar to the respective states with $v_b = 0$
and can again be analyzed in the $(R_1,R_2)$ plane. Bending excitation
introduces an additional \lq layer' of nodes and enhances mixing
between the combination states $(v_\phi,0,1)$ and $(v_\phi-4,1,1)$ and
both assignments are given in some panels.  
Above 78\,000\,cm$^{-1}$, the assignments become even less
unique. Several factors contribute: Resonances substantially
broaden, states with $v_b > 1$ are found, and mixing among the three
vibrtaional motions increases.
Conservative assignment associates the simple looking
high energy bands with progressions $(v_\phi,0,1)/(v_\phi-4,1,1)$ and
$(v_\phi,1,0)$, although this selection is by no means unique. 

The shape of the lower adiabatic potential in the plane of the tuning
and the coupling modes of a CI is often referred to as  
a Mexican hat, especially in the context of the $E\times\epsilon$
Jahn-Teller effect.\cite{KDC84}  
A closed CI seam discussed in this paper adds another hat
to the rack: A \lq Mongolian hat', sketched in Fig.\ \ref{wf01}, 
is a piece of headwear which describes best the shape of the upper adiabatic PES
in the $(R_1,R_2)$ plane containing the CI seam. The hindered
rotation excitations are winding along the hat brim (the CI seam), while
consecutive radial excitations extend towards (and eventually engulf)
the top of the hat. 

The closed CI seam leaves a clear mark in the calculated
absorption spectrum. Frequencies in the progression of pure $v_\phi$
excitations are shown in Fig.\ \ref{specmod}, in panel (b) for the full
calculation and in panel (c) for the 2D model. Consider the model
first. Energy intervals grow between $E_{\rm ph}=73\,000$\,cm$^{-1}$ and
76\,000\,cm$^{-1}$, and this is consistent with 
a pseudorotational motion around
the seam and the associated quantization with $E_\phi \sim
v_\phi^2$. At the same time, the frequency in the 
$(v_\phi,0)_{\rm 2D}$ progression is not monotonic, and 
there are two dips at low and high energies. 
Minimum in a progression frequency is often
an indication that a critical energy in the potential (a barrier top or a
saddle)\cite{CRS87} or in the phase space (a
separatrix) has been reached.\cite{K95A,WHGDSCHKSF00} Possible sources of
the rise-and-fall pattern in the 2D model are illustrated in the lower
panels of Fig.\ \ref{specmod} showing 
one-dimensional cuts through the PES of the
adiabatic $3^1\!A'$ state, taken along the seam [panel (d)] and
across the seam [panel (e)]. Angular
coordinate $\phi$ along the seam in (d) is chosen such that $\phi =
180^\circ$ for $R_1 = R_2 = 2.24\,a_0$;  
the \lq radial' coordinate in (e) runs along
the antisymmetric stretch $R_-$, and the origin 
$R_- = 0$ is the point $R_1 = R_2 = 2.5\,a_0$. The potential along the
seam has two global minima separated by a local one at $\phi =
180^\circ$. The first frequency dip (gray color) 
in the $(v_\phi,0)_{\rm 2D}$ progression involves the excited 
state in the global well and the ground state in the local well. 
The second frequency dip (yellow color) involves the states located
near two types of critical points: 
The barriers, separating the local and the global minima along the seam
[$\phi=135^\circ$ and $225^\circ$ in panel (d)], and the 
top of the Mongolian hat, separating the global minima 
across the seam [$R_- = 0$ in panel (e)]. 

The principal topographic features, captured by the 2D model, shape the
spectrum in the full calcultion, too. The frequency in progression
$(v_\phi,0,0)$ in Fig.\ \ref{specmod}(b), although numerically
different from the 2D case, grows steadily up to $E_{\rm ph} \approx 
75\,000$\,cm$^{-1}$ reflecting consecutive 
pseudorotational excitations along the CI seam. The frequency dip at
high energies indicates that the top of the Mongolian hat has been
reached, while the low energy dip is absent because 
the origin of the progression lies above the bottom of the local minimum at
$\phi=180^\circ$.  

The accuracy of the ab initio PESs does not allow a line by line
comparison with the measured spectrum. To which extent
the above assignment is applicable to the experiment? 
Energy intervals in the major progression  measured below 
$E_{\rm ph}=78\,000$\,cm$^{-1}$ 
are shown in Fig.\ \ref{specmod}(a), and the  rise-and-fall
pattern,  inherent to the calculated spectra, 
appears to be distinctly recognizable, including a stretch of 
pseudorotational excitations and dips impying potential saddles. This
observation should not be overinterpreted 
--- shown is one progression, and this progression is
rather short. If, however, the agreement with the calculations is 
more than a mere coincidence, the 133\,nm band
would be the first documented spectroscopic observation of a closed
CI seam. 

I would like to conclude the discussion of the 133\,nm band by
pointing out an alternative explanation for the double peaked lines
above 78\,000\,cm$^{-1}$. Their assignment in terms of  
 progressions $(v_\phi,0,1)$  and $(v_\phi,1,0)$ has been given only 
for the
parallel component of the total spectrum.
Decomposition into $\sigma_\parallel$ and $\sigma_\perp$ 
at the bottom of Fig.\ \ref{spec2}(b) demonstrates
that intensities in the two transitions become
comparable at high photon energies. In fact, each double peaked line
in the total spectrum consists of a low energy parallel peak (green line)
overlapped with a high energy perpendicular peak (red line). 

\subsection{Vibronic assignments of the low energy band} 
\label{vibassl}

Assignment of the 148\,nm band is a challenging problem:
Diffuse structures are weak, irregular, and overlapping.  Substantial bending
excitation has been suggested for this 
band\cite{SFCCRWB92,YESPIM96} originating from states 
$2^1\!A'$ and $1^1\!A''$ with 
strongly bent equilibrium geometries.\cite{D63,SFCCRWB92} 
The present calculations demonstrate that actual molecular
motion is difficult to define in terms of the familiar
normal modes: Energy above 
equilibrium is high and the anharmonic and vibronic couplings are
substantial. The results summarized in this section at best provide an 
initial point from which a genuine analysis could be undertaken.

In the experimental spectrum, almost every diffuse structure 
below 72\,000\,cm$^{-1}$ comprises several transitions visible as a
group of peaks or as a shoulder [Fig.\ \ref{spec1}(b)]. 
Diffuse structures in the calculated spectrum look 
similar, especially after thermal averaging,
although irregularities in the
energy intervals, widths, and intensities preclude a meaningful
matching with experiment. In order to enable comparison,
line subsets are selected in both spectra. In the experimental spectrum, 
centers of the diffuse peaks with the largest intensities are chosen
as indicated in Fig.\ \ref{spec2}(a). In the calculated spectrum, 
only the parallel
transition $(J_i=0,\Omega_i=0) \rightarrow (J_f=1,\Omega_f=0)$
is initially considered, and resonance states ${\bm
  \Psi}_n$ with the largest intensities are taken to represent 
diffuse peaks in $\sigma_\parallel$ [Fig.\ \ref{spec2}(d)]. 

Energy intervals between the selected peaks are 
shown in Fig.\ \ref{freqleb} ($E_{\rm ph} \ge
60\,000$\,cm$^{-1}$). Experimental peaks are spaced by 
400\,cm$^{-1}$---$700$\,cm$^{-1}$. Spacings between the intense
resonances vary in similar limits although  the agreement is only
qualitative. Moreover, visual assignment of the wave functions fails
for the selected resonances. A typical example is given 
in Fig.\ \ref{resleb} which depicts the $2A'$ adiabatic
component of a state lying at $E_{\rm ph} = 8.38$\,eV, 
i.e. $\sim 3$\,eV above equilibrium of the $2A'$ state 
and  $\sim 1$\,eV above dissociation threshold. The large 
excess energy is the reason for the bizzar 
nodal patterns of the $2A'$ resonance components, and ---
although wave functions of different states  
in Fig.\ \ref{spec2}(d) have
different nodal patterns --- their complexity is beyond 
rationalization in terms of progressions. 

An alternative \lq bottom-up' approach in this case is to study the spectrum of
the bent $2^1\!A'$ state from the ground state upwards and and to follow the
characteritic frequencies
with growing internal energy. Bent electronic states have $C_{2v}$ equilibria
(see Table III of paper I), and the relevant 
vibrational quantum numbers are the symmetric stretch $v_s$, the bend
$v_b$, and the antisymmetric stretch $v_a$. All wave functions 
in the full calculations, both resonance and bound,  were
screened in order to find  states, $2A'$ adiabatic components of which 
represent pure excitations in the three vibrational modes. 
Examples of pure stretching and bending wave functions 
are shown in  Figs.\ \ref{purprog1}  and \ref{purprog2}, respectively.
The progression $(v_s,0,0)$ is the simplest: Consecutive wave functions  
have parallel nodal lines (normal to which indicates the 
normal mode direction), and the
symmetric stretch frequency $\omega_s \approx 1400$\,cm$^{-1}$ is
mildly anharmonic and depends smoothly on energy (see Fig.\ \ref{freqleb}, 
$E_{\rm ph} < 60\,000$\,cm$^{-1}$). In the progression
$(0,0,v_a)$, only states with even $v_a$ are visible, because the 
transition is parallel and both $\mu_y$ and $\mu_z$ are symmetric with
respect to interchange of $R_1$ and $R_2$. The antisymmetric stretch
frequency, $\omega_a \approx 900$\,cm$^{-1}$, grows
at low energies, passes through a maximum near $49\,000$\,cm$^{-1}$
and then decreases (Fig.\ \ref{freqleb}); 
nodal lines in the lower panels of Fig.\ \ref{purprog1}
noticably bend in the $(R_1,R_2)$ plane with growing $v_a$. The
anharmonicity is the strongest in the bending progression
$(0,v_b,0)$. While the direction of the bending vibration is rather
well preserved, its frequency in Fig.\ \ref{freqleb} varies from
$\omega_b(A') \approx 550$\,cm$^{-1}$ down to 380\,cm$^{-1}$, experiencing
a slow rise followed by a rapid decline above 49\,000\,cm$^{-1}$. 
The mode coupling and the density of states grow
substantially above $E_{\rm ph} = 52\,000$\,cm$^{-1}$, 
rendering further search for pure excitations unfeasible. This leaves
a 10\,000\,cm$^{-1}$ wide energy gap between the last assigned state
and the resolvable red edge of the 148\,nm band.
In view of the capricious anharmonicities, 
an extrapolation over this gap would seem unreliable
for any mode but the symmetric stretch. However, if it is granted that 
the two lowest frequencies, $\omega_b \sim 350$\,cm$^{-1}$ and 
$\omega_a < 900$\,cm$^{-1}$, keep their values up to and above 
60\,000\,cm$^{-1}$, one would expect the diffuse peaks to reflect
strongly excited bending and antisymmetric stretch motions. 

One major drawback of the above discussion is the missing experimental
counterparts of the calculated vibrational frequencies: 
The measured intensities are low and the rotational structure is unresolved.  
Truly exceptional in this respect is the high resolution 
study of the wavelength range 172\,nm---198\,nm, performed by
Cossart-Magos and co-workers\cite{CLP92} who were able to detect many
absorption bands with rotational structure and to 
isolate two vibrational progressions between 51\,000\,cm$^{-1}$ and 
57\,000\,cm$^{-1}$. This energy range overlaps with the energies in
which pure excitations in the quantum mechanical calculations still could
be detected.  The
progressions were assigned to the electronic state $1^1\!A_2$, i.e. the
bent state $1^1\!A''$, 
excited in a perpendicular transition from $\tilde{X}$. Frequencies
in the experimental progressions are shown with open
triangles and squares in Fig.\ \ref{freqleb}; open circles are the
frequency shifts between states belonging different
progressions. In Ref.\ \onlinecite{CLP92}, both progressions were
interpreted as consecutive bending excitations differing
by one quantum of the symmetric stretch, giving experimental estimates
of $\omega_b \approx 600$\,cm$^{-1}$ and $\omega_s \approx
1370$\,cm$^{-1}$. Relevant theoretical progressions are 
shown in Fig.\ \ref{freqleb} with red symbols. They are obtained by
calculating eigenstates in the perpendicular transition 
$(J_i=0,\Omega_i=0) \rightarrow (J_f=1,\Omega_f=1)$ in many energy
windows and looking for the wave functions ${\bm  \Psi}_n$, 
$1A''$ adiabatic components of which represent consecutive bending excitations. 
Because $\mu_x(R_1,R_2) = -\mu_x(R_2,R_1)$, all excited states involve  
odd number of quanta in the antisymmetric stretching mode $v_a$. 
The progression shown in the lower part of 
Fig.\ \ref{freqleb} (red squares) 
is $(1,v_b,1)$, and the frequency shifts with
respect to the progression $(0,v_b,1)$ are given in the upper part of 
Fig.\ \ref{freqleb} (red circles). The calculation clearly confirms
the experimental assignment. Agreement with the experiment is
quantitative, especially for the bending frequency $\omega_b(A'')$
in the $(1,v_b,1)$
progression, which lies within 10\,cm$^{-1}$ of the experimental
values. Somewhat less accurate is the symmetric stretch frequency
shift and its energy variation. Most probably, this implies that the  
observed progressions are built on higher 
symmetric stretch excitations than $v_s = 0$ and $v_s = 1$. The
calculated intensity generally grows with $v_s$, but sufficiently long bending
progressions with $v_s \ge 2$ become tough to identify. 

The bending progressions in the parallel and perpendicular
transitions shown in in Fig.\ \ref{freqleb} 
are strikingly different: $\omega_b(A'')$ in $\sigma_\perp$ 
smoothly depends on energy and is only slightly anharmonic, while  
$\omega_b(A')$ in $\sigma_\parallel$  has a strong non-monotonic energy
dependence. This is surprising because the 
equilibria of the bent states $2^1\!A'$ and $1^1\!A''$ are similar.
The sole difference is in the equilibrium angle, 127$^\circ$ in 
$1^1\!A''$ vs. 118$^\circ$ in $2^1\!A'$, but it can hardly be made
responsible for the observed discrepancy in the bending frequencies. 
Instead, the strong energy dependence of $\omega_b(A')$ 
stems from another property of the $2^1\!A'$ PES along the bending
angle: In the calculation, it is this state to which the carbene-type bent OCO
minimum at $70^\circ$
belongs after local diabatization of bent CIs. This minimum at
$\alpha_{\rm OCO} = 73^\circ$ is is
clearly seen in the contour maps in Fig.\ \ref{purprog2}. Near 
$E_{\rm ph} \approx48\,000$\,cm$^{-1}$, the ground vibrational state
appears in this local OCO minimum shown in the left lower panel in Fig.\ 
\ref{purprog2}. Above $50\,000$\,cm$^{-1}$, the bending states in the
local and in the global
equilibria become connected via a saddle point, the main bending
progression includes states partially delocalized between the two
wells [right lower panel in Fig.\  \ref{purprog2}], the configuration
space available to bending vibrations suddenly expands, 
and it is this energy range in which $\omega_b(A')$ starts to decrease. In
contrast, the $1^1\!A''$ state, diabatized at bent CIs in a different
symmetry block, loses the OCO bent minimum which becomes reassigned to state
$3^1\!A''$. As a consequence, the energy dependence of
$\omega_b(A'')$ is monotonic and inconspicuous. 

The difference in bending progressions in the states 
$2^1\!A'$ and $1^1\!A''$ and its connection to the local OCO minimum
along $\alpha_{\rm OCO}$ allows one to explain  
the difference in the diffuse structures in the parallel and
perpendicular excitations of the 148\,nm band in Fig.\ \ref{spec2}(b). 
A nearly constant frequency $\omega_b(A'')$ correlates with the 
regular progression of peaks in $\sigma_\perp$. On the other hand, the
bending frequency $\omega_b(A')$ in the parallel excitation is small
and strongly energy dependent --- and the diffuse peaks in
$\sigma_\parallel$ are narrowly spaced, irregular, and
overlapping. Simultaneous analysis of both types of transitions
indirectly confirms that the observed absorption lines are indeed due
to strong bending excitations: Qualitatively different bending
potentials support qualitatively different spectra. 
Moreover, the \lq congestion' in
the 148\,nm parallel band is interpreted as a direct consequence 
of the carbene-type OCO \lq cyclic' minimum in the bent $2^1\!A'$ state. 

It is worth reminding the reader that the question of electronic
assignment of the carbene-type OCO goes beyond the diabatic states $2^1\!A'$,
$1^1\!A''$, or the ground electronic state in which it was discovered 
by Xantheas and Ruedenberg.\cite{XR94} In paper I, the origin of this
minimum was traced to the states $4^1\!A'$ and $4^1\!A''$ correlating
at linearity with the Rydberg state $2^1\Delta_u$. The influence of
the carbene-type OCO on the diffuse 148\,nm band, uncovered in the
above calculations, suggests that a systematic analysis of this band
has to
include the first four $A'$ and $A''$ states properly diabatized and fully
interacting at multiple bent CIs.

\subsection{Vibrational assignments in the state $^1\Sigma_u^-$} 
\label{vibasssi}

The spectrum of the state $2^1\!A''$ ($^1\Sigma_u^-$ at linearity) 
is shown in Fig.\ \ref{spec3}. 
It stretches from 65\,000\,cm$^{-1}$ 
to about 83\,000\,cm$^{-1}$ and consists of several series of 
narrow peaks. All vibrational states, excited in this transition, 
lie below $92\,430$\,cm$^{-1}$, i.e. the threshold energy of
the channel ${\rm O}(^3\!P) + {\rm  CO}(a^3\Pi)$
with which the state  $^1\Sigma_u^-$ correlates. 
Predissociation is purely non-adiabatic and therefore inhibited in these
calculations. Because of the extremely small TDM $\mu_x$, the
intensity of this band, excited in a perpendicular transition from $\tilde{X}$, 
does not exceed $6\times 10^{-20}$\,cm$^{-2}$. The 
frequencies of the bending, symmetric and antisymmetric stretching
modes are $\omega_b =576.7$\,cm$^{-1}$, 
$\omega_s =1015.5$\,cm$^{-1}$, and $\omega_a =1118.7$\,cm$^{-1}$,
respectively. The vibrational ground state $(0,0,0)$ lies at 
64220\,cm$^{-1}$ but remains invisible in a perpendicular transition
in which only states with odd number of antisymmetric stretch quanta $v_a$
are excited.
Most intense lines in the spectrum belong to one of the three main 
progressions, assigned by visual inspection of the corresponding wave
functions, although excitations are 
easily predictable from the FC geometries. The state $^1\Sigma_u^-$
is linear, the angular dependence of the TDM $\mu_x$ is weak, and the
bending vibration remains unexcited. In contrast, the equilibrium
bond lengths, $R_1 = R_2 = 2.4\,a_0$, are substantially extended as
compared to the equilibrium in $\tilde{X}$ at $R_1 = R_2 =
2.20\,a_0$, with the implication that the spectrum is built on stretching
excitations. The nodal patterns of the vibrational 
eigenfunctions clearly demonstrate that the three main 
progressions, $(v_s,0,1)$, $(v_s,0,3)$, and $(v_s,0,5)$,  involve 
consecutive excitations of the symmetric stretch mode 
accompanied by one, three, or five quanta of the antisymmetric
stretch. Maximum intensity in each progression is reached for the same
$v_s = 7$, and the intensity distributions around the maximum
are approximately bell shaped.

\section{Summary and concluding remarks}
\label{concl}

This paper analyzes the absorption spectrum of carbon dioxide in the
wavelength range 120\,nm --- 160\,nm by means of quantum mechanical
calculations using PESs of five singlet valence electronic states
and their coordinate dependent TDM vectors with the ground electronic
state, constructed as described in paper I.  
The main results are as follows: 
\begin{enumerate}
\item The UV spectrum of five valence states, calculated for $T=190$\,K via
Boltzmann averaging of optical transitions from many initial rotational
states, reproduces the experimental two-band 
spectral envelope, the positions of the
absorption maxima, their FWHMs, peak intensities, 
and frequencies of diffuse structures in each band. 
\item The two absorption bands have been assigned at several levels of
  graining. Electronic assignment, reflecting the strong HT effect
  and influenced by the vibronic
  interactions at CIs, is given in the adiabatic
  representation (low energy band: bent states $2^1\!A'/1^1\!A''$; 
high energy band: linear states $3^1\!A'/3^1\!A''$; 
gap between the two bands: state $2^1\!A''$).  
The parallel and perpendicular components of the full spectrum are
isolated. The parallel transition, the RT coupling in which is suppressed
for excitations from the ground vibrational state in $\tilde{X}$,  
is shown to be dominant for all but the highest photon energies.  
Finally, individual diffuse structures are 
assigned using multicomponent wave functions of metastable resonance states.
\item  
{\it In the high energy band}, the 
main progressions correspond to consecutive excitations of the
pseudorotational motion around the closed loop of the CI
seam; progressions differ in the number of nodes along the radial mode
perpendicular to the closed seam. Systematic frequency growth in the
pseudorotational  progressions is interpreted as a spectroscopic
hallmark of the closed seam. 
\item {\it In the low energy band}, the diffuse peaks, 
  associated with bending excitations in the states
  $2^1\!A'$ and $1^1\!A''$,  are analyzed in both
parallel and perpendicular transitions. Calculations
  reveal a dramatic difference in the spectra of the
  two states. The perpendicular band (mainly $1^1\!A''$), 
  bending progressions in which are nearly equidistant, is rather
  regular. In contrast, bending excitations in
the parallel band (mainly $2^1\!A'$), 
are very anharmonic and irregular. This difference
is interpreted as a manifestation of the 
carbene-type \lq cyclic' OCO minimum in the PES of the 
$2^1\!A'$ state.
\end{enumerate}

The present study focuses on spectroscopic aspects of the UV
photodynamics of CO$_2$. Mechanisms of photodissociation are
discussed only insofar as they become apparent 
in the analysis of the absorption spectrum. This piece
will be completed in the discussion of photofragment distributions
which will be published separately. However, preliminary conclusions
can be deduced from the density distributions of resonance wave
functions ${\bm \Psi}_n$
calculated and discussed above. Of interest here are the 
$2A'$ components of ${\bm \Psi}_n$ as depicted for example in
Fig.\ \ref{resleb}. Although useless in establishing the
vibrational assignments, they
 rather clearly illustrate the dissociation path taken by the
 molecule once it arrives in a bent adiabatic state. Because of 
 CIs in linear CO$_2$, the inner regions of 
states $2^1\!A'$ and  $1^1\!A''$ are separated from the
dissociation asymptotes by barriers. 
On the other hand, the $2^1\!A'$ and  $1^1\!A''$ 
potentials along $\alpha_{\rm OCO}$
  in the direction of global bent equilibria are barrierless. This
  directional anisotropy is reflected in the probability density in
  the right panel of Fig.\
\ref{resleb}: The wave function traces out a path leading from the FC region
to smaller angles; 
hardly any probability is found along the linear dissociation path. 
Substantial extension of the CO bond, seen in  the left panel of Fig.\
\ref{resleb}, occurs predominantly in bent molecular
configurations with $\alpha_{\rm OCO} < 140^\circ$. This shape is
representative of the $2A'$ or $1A''$ components of ${\bm \Psi}_n$
both below the CI energy (where the bent states are directly
accessed) and above the CI energy (where they become populated
indirectly via the CI funnels at linearity). 

%In conclusion, I would like to comment on the relation between the
%excitation of the pseudorotational motion along the CI seam loop and
%the standard Jahn-Teller effect. The two band shape of the absorption
%spectrum of CO$_2$ is reminicent of the photoelectroni spectra which
%are routibnely observed in JT active molecules... ?

%%%%%%%%%%%%%%%%%%%%%%%%%%%%%%%%%%%%%%%%%%%

%\newpage

\acknowledgments{ 
Financial support by the  Deutsche
  Forschungsgemeinschaft is gratefully acknowledged
}
\clearpage

%\bibliography{sergy}

\begin{thebibliography}{10}
\bibitem{NOTE-CO24-1}
{\rm Ab initio calculations in paper I use a different coordinate system
  $(x',y',z')$: $x'$ is orthogonal to the molecular plane, $z'$ runs along one
  of the CO bonds and $y' \perp z'$. It is related to the principal axes system
  by an orthogonal transformation.}
\bibitem{NOTE-CO21-4}
{\rm \lq PolyWave' is a package of FORTRAN90 programs for iterative quantum
  mechanical calculations of bound states, dissociative resonance states,
  photoabsorption or photoemission spectra, as well as product state
  distributions in molecules with up to six internal degrees of freedom and $N
  \le 9$ coupled electronic states. The package is available from the author
  upon request.}
\end{thebibliography}

\newpage
\clearpage

\begin{figure}[t]
\caption{(a) Experimental absorption cross
section of CO$_2$ ($T=195$\,K; in cm$^{2}$) as a function of photon energy (in
  10$^3$\,cm$^{-1}$). Numbers on the cross section axis denote powers
  of ten. The photon energy range studied in this work is indicated
  with two dashed lines. (b) Absorption spectrum, calculated at
  $T=200$\,K (blue line, vertically shifted by $0.9\cdot10^{-18}$\,cm$^{2}$), 
compared with the experimental spectrum
  (black line).   Electronic assignment in terms of adiabatic
  states is given. Experimental data are from Ref.\ \onlinecite{YESPIM96}. 
Theoretical spectrum is shifted by 400\,cm$^{-1}$
  to lower energies. 
\label{spec1}
}
\end{figure}

\begin{figure}[t]
\caption{Cuts through the PESs of the adiabatic
  states along (a) OCO angle (fixed coordinates are $R_{1} =
  2.2\,a_0$ and $R_{2} = 2.3\,a_0$) and (b) one CO bond distance
  (fixed coordinates are $R_{\rm CO} = 2.3\,a_0$ and $\alpha_{\rm OCO} =
  175^\circ$). The thick black line in both panels is the ground
  electronic state $\tilde{X}$; excited
  $A'$ and $A''$ states are shown with thin black and brown
  lines, respectively. Red lines and arrows illustrate 
the non-vertical excitation process and 
the effect of the TDM on the initial vibrational wave function
$\psi_0$ in $\tilde{X}$. $\psi_0$ is shown in both
  panels with a dashed line, and the shapes of the 
initial state in the optical transitions via $y$  and $x$ components
of the TDM, $\mu_y\psi_0$ (a)  and $\mu_x\psi_0$ (b) are shown with solid 
lines. Arrows, indicating excitations, 
pass through the maximum of $\psi_0$ and the maxima of
$\mu_y\psi_0$  and $\mu_x\psi_0$. 
\label{copot1}
}
\end{figure}

\begin{figure}[t]
\caption{Spatial components $\mu_x$ (a,d), $\mu_y$ (b,e), and $\mu_z$
(c,f)  of the ab initio TDM vectors for the adiabatic states 
$2^1\!A'$ (purple), $3^1\!A'$ (brown), $1^1\!A''$ (black), $2^1\!A''$ (purple), 
and $3^1\!A''$ (brown) from the ground state $\tilde{X}^1\!A'$. 
$A'$ states are shown with solid, $A''$ states with open symbols. 
Dependences on the CO bond distance are shown in (a)---(c) [fixed
coordinates are $R_2 = 2.2\,a_0$ and $\alpha_{\rm OCO}=179^\circ$] and
dependences on the OCO angle are shown in
(d)---(f) [fixed coordinates are $R_1 = 2.2\,a_0$ and $R_2 = 2.3\,a_0$]. 
TDMs are measured in atomic units. The coordinate axes are sketched in (b): 
$y$ is directed along the molecular figure axis, and
$z$ is normal to $y$ in the molecular plane, and $x$ is normal to the
molecular plane. 
\label{tdms}
}
\end{figure}

\begin{figure}[t]
\caption{(a) The absorption spectrum  $\sigma(E_{\rm ph}|J_i=0)$ for a
  single initial state $J_i = 0$ (blue) compared with
  the spectrum of Ref.\ \onlinecite{YESPIM96} measured at
$T=195$\,K (black). 
Thin sticks indicate the major
experimental peaks separated into a strong (brown) and a weak
(blue) progression in the high energy band, and a strong progression in
the low energy band (black). Their positions 
are taken from Table X of Ref.\ \onlinecite{RMSM71} 
augmented with data from Ref.\ \onlinecite{YESPIM96}. 
Panels (b-d) show decomposition of the calculated
spectrum and vibronic assignments: 
(b) $\sigma(E_{\rm ph}|J_i=0)$ spectrum
(thin gray line) broken down into two low resolution components 
  stemming from excitations of 
the bent ($2^1\!A'$, $1^1\!A''$, thick black line) and
  linear ($3^1\!A'$, $3^1\!A''$, thick blue line) adiabatic states.  
 Two spectra at the bottom are the parallel
 ($\sigma_\parallel$,  thick green line) 
and the perpendicular ($\sigma_\perp$,  thick red line) components of 
$\sigma(E_{\rm ph}|J_i=0)$; (c) Decomposition of $\sigma_\parallel$
(green) in terms of resonance
  states of the Hamiltonian Eq.\ (\ref{ham02}). Gray sticks mark 
  resonance positions (every 3d calculated
  state is shown). Black line is the sum of Lorentzians in Eq.\
  (\ref{sigmares}); 
  (d) Vibronic assignment of $\sigma_\parallel$
 (green) in terms of 
the most  intense resonances shown at the bottom of the
  panel with thin sticks. Vertical shift of the calculated
  spectra is $0.9\cdot10^{-18}$\,cm$^{2}$ in panels (a) and (b), and 
$0.6\cdot10^{-18}$\,cm$^{2}$ in  panels (c) and (d).  
\label{spec2}
}
\end{figure}

\begin{figure}[t]
\caption{{\it Left and right columns } depict
$3A'$ adiabatic components of the 3D wave functions of resonance
  states in progressions $(v_\phi,0,0)$ and
  $(v_\phi,0,1)$, respectively. 
Shown in molecular coordinates is one particular contour
    $\left|\Psi(R_1,R_2,\alpha_{\rm OCO})\right|^2 = {\rm const}$.
 The contour is
    viewed along the $\alpha_{\rm OCO}$ axis and appears \lq
    projected' onto the $(R_1,R_2)$ plane. Shading emphasizes the 3D character
    of the plots. {\it Middle column}
  depicts contour maps of the eigenstates $\psi_{\rm 2D}(R_1,R_2)$
  calculated in the 2D single state model.  All states belong to 
  progression $(v_\phi,0)_{\rm 2D}$; red contours represent
  positive, blue contours negative values of $\psi_{\rm 2D}$.  
In all panels, a 2D contour map of the $3A'$
adiabatic potential for $\alpha_{\rm OCO} = 179^\circ$ (gray lines) 
and the CI seam (black thick line) are sketched. Axis tic labels are
in $a_0$. States marked $(v_\phi,0)_{\rm 2D}^\star$ are mixed 
with states $(v_\phi-4,1)_{\rm 2D}^\star$
shown in Fig.\  \protect\ref{wf02}.  
The top middle panel shows author's drawing of a Mongolian hat.
\label{wf01}
}
\end{figure}

\begin{figure}[t]
\caption{{\it Upper panels}: 
$3A'$ adiabatic components of the 3D wave functions of resonance
  states in the progression $(v_\phi,1,0)$. The state
  $(3,1,0)$ could not be found. {\it Lower panels }    
  depict contour maps of the eigenstates $\psi_{\rm 2D}(R_1,R_2)$
  calculated in the 2D single state model and belonging to the 
  progression $(v_\phi,1)_{\rm 2D}$. 
The layout of all wave functions is the same as in 
Fig.\  \protect\ref{wf01}. The 2D states are mixed with states  
$(v_\phi+4,0)_{\rm 2D}$; counterparts of those marked 
$(v_\phi,1)_{\rm 2D}^\star$ are shown in 
 Fig.\  \protect\ref{wf01}. 
\label{wf02}
}
\end{figure}

\begin{figure}[t]
\caption{{\it Upper panels}: 
Frequencies (a) in the major exprimental progression, (b) in the
  progression $(v_\phi,0,0)$ in the full calculations,
  and (c) in the progression $(v_\phi,0)_{\rm 2D}$ in the 2D model.  
In (b), frequencies are lifted by
  50\,cm$^{-1}$; in (c), the frequency scale is omitted. 
  Arrows emphasize the positions of dips in the progressions. {\it Lower
    panels}: Potential of the $3^1\!A'$ state (d) at $\alpha_{\rm
    OCO}=179^\circ$ 
along the ab initio CI seam and (e) across the Mongolian hat top. 
Angular
coordinate $\phi$ along the seam in (d) is chosen such that $\phi =
180^\circ$ for $R_1 = R_2 = 2.24\,a_0$;  
the \lq radial' coordinate in (e) runs along
the antisymmetric stretch $R_-$, and the origin 
$R_- = 0$ is the point $R_1 = R_2 = 2.5\,a_0$. 
Potential curves are rised through the
zero point energies of \lq missing' coordinates, $\hbar\omega_r/2$
and $\hbar\omega_\phi/2$ in (d) and (e), respectively. 
Vibrational ladder in the 2D model is indicated in (d) and (e).
Energy spacings forming two dips in the progression in panel (c) are
highlighted with gray and yellow. 
\label{specmod}
}
\end{figure}

\begin{figure}[t]
\caption{Shown left of the vertical dashed line are frequencies
  in the pure progressions of eigenstates ${\bm  \Psi}_n$ in which the 
$2^1\!A'$ adiabatic component (purple, blue, and brown lines and symbols)
or the $1^1\!A''$ adiabatic component (red lines and symbols)
are most populated. Open black symbols are the two experimental vibrational
  progressions (triangles and squares), as well as the 
  frequency shift between their band maxima (circles) taken from
  Table 3 of Ref.\ \onlinecite{CLP92}. 
Shown right of the vertical dashed line are energy
  intervals between adjacent diffuse peaks in the experimental
  spectrum of Ref.\ \onlinecite{YESPIM96}
(black open circles) and between most intense
  resonance states (green solid circles). Note that the frequency
  scale changes in the upper part of the figure. 
\label{freqleb}
}
\end{figure}

\begin{figure}[t]
\caption{$2A'$ adiabatic component of the
  resonance state with $E_0 = 8.3795$\,eV and $\Gamma_0 =
  62$\,cm$^{-1}$. Shown is one particular contour
    $\left|\Psi(R_1,R_2,\alpha_{\rm OCO})\right|^2 = {\rm const.}$, with
    the same constant in both panels. The contour is
    viewed  along $\alpha_{\rm OCO}$ (left panel) and
  along the CO bond (right panel). Shading emphasizes the 3D character
    of the plots. The 2D contour maps of the
  adiabatic state $2^1\!A'$ in the $(R_1,R_2)$ plane and in the 
$(\alpha_{\rm OCO},R_2)$ plane are sketched in the left and right
panels, respectively.     
Thick black lines trace out the probability density buildup away from 
the linear FC region. 
\label{resleb}
}
\end{figure}

\begin{figure}[t]
\caption{$2A'$ adiabatic components of the 3D wave functions of states 
  in the progression $(v_s,0,0)$ (upper panels)  and $(0,0,v_a)$
  (lower panels). The layout of all wave functions is the same as in 
  the left panel of Fig.\  \protect\ref{resleb}.
\label{purprog1}
}
\end{figure}

\begin{figure}[t]
\caption{$2A'$ adiabatic components of the 3D wave functions of states 
  in the progression $(0,v_b,0)$.  In the left lower frame, the ground 
vibrational state in the carbene-type OCO minimum is shown and labeled 
$(0,0,0)_c$. The layout of all wave functions is
  the same as in the right panel of Fig.\  \protect\ref{resleb}.
\label{purprog2}
}
\end{figure}

\begin{figure}[t]
\caption{Absorption spectrum of the state $^1\Sigma_u^-$ and assignments of the
three progressions $(v_s,0,v_a)$ in which the most intense absorption
lines are found. 
\label{spec3}
}
\end{figure}

%%%%%%%%%%%%%%%%%%%%%%%%%%%%%%%%%%%%%%%%%%%%%%%%%%%%%%%%%%%%%%%%%%%%%%%%%%%%
%              THE PART BELOW CONATINS ONLY FIGURES                        %
%%%%%%%%%%%%%%%%%%%%%%%%%%%%%%%%%%%%%%%%%%%%%%%%%%%%%%%%%%%%%%%%%%%%%%%%%%%%

\clearpage
\newpage
\mbox{ }
\vspace{1cm}

\includegraphics[angle=0,scale=0.7]{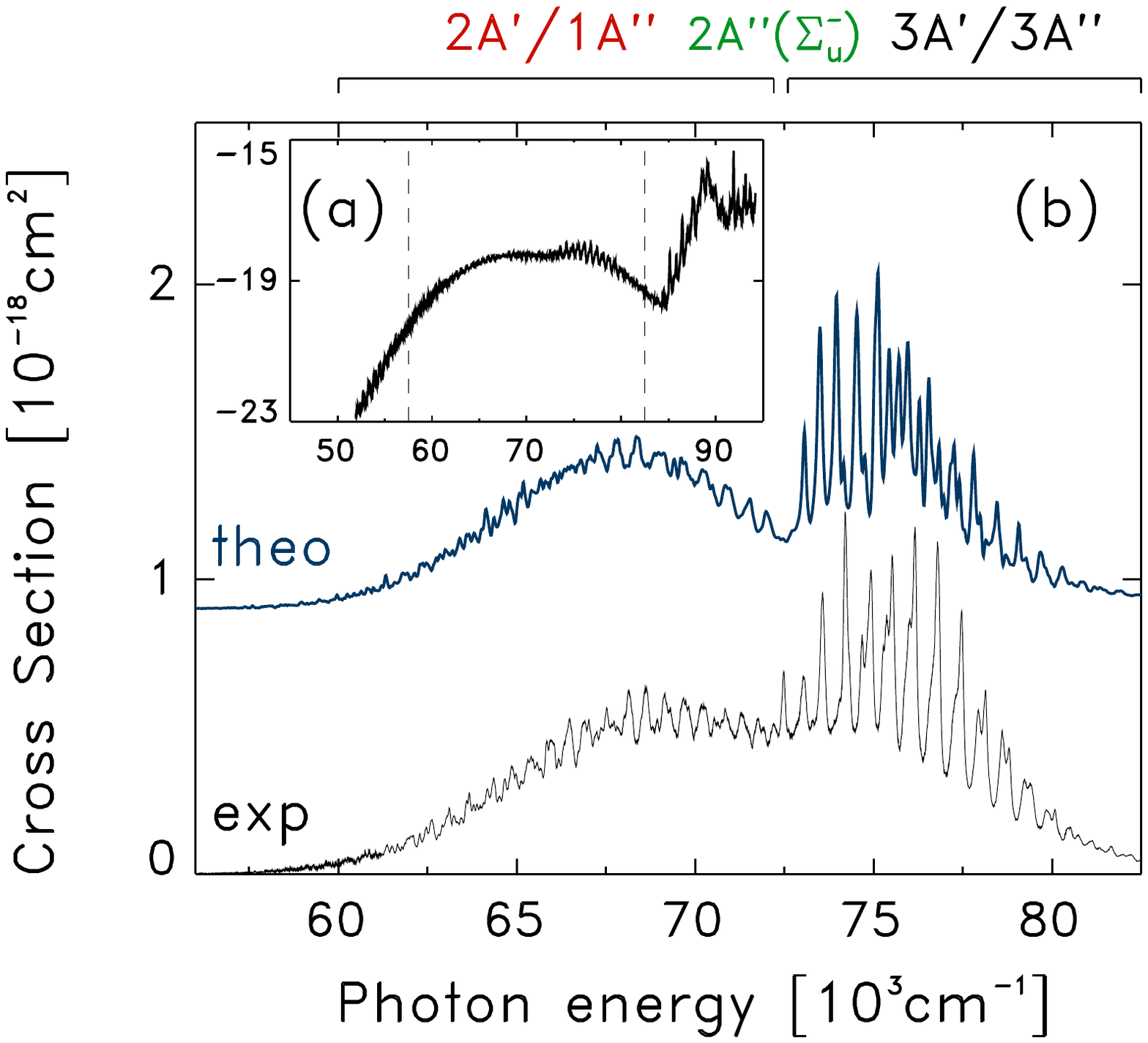}

\vspace{-1cm}

Fig.\ 1

\newpage
\mbox{ }
\vspace{1cm}

\includegraphics[angle=90,scale=0.5]{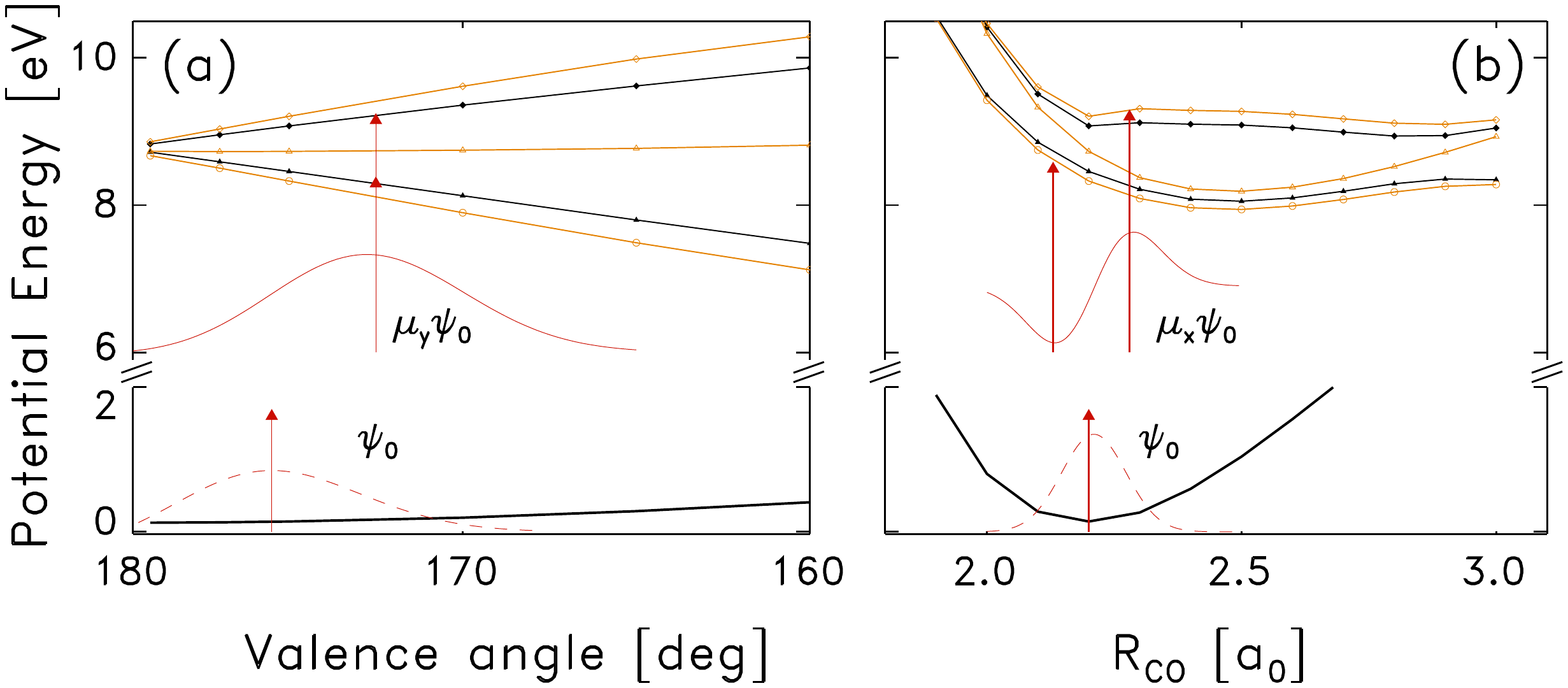}

\vspace{2cm}

Fig.\ 2

\newpage
\mbox{ }
\vspace{1cm}

\includegraphics[angle=0,scale=0.7]{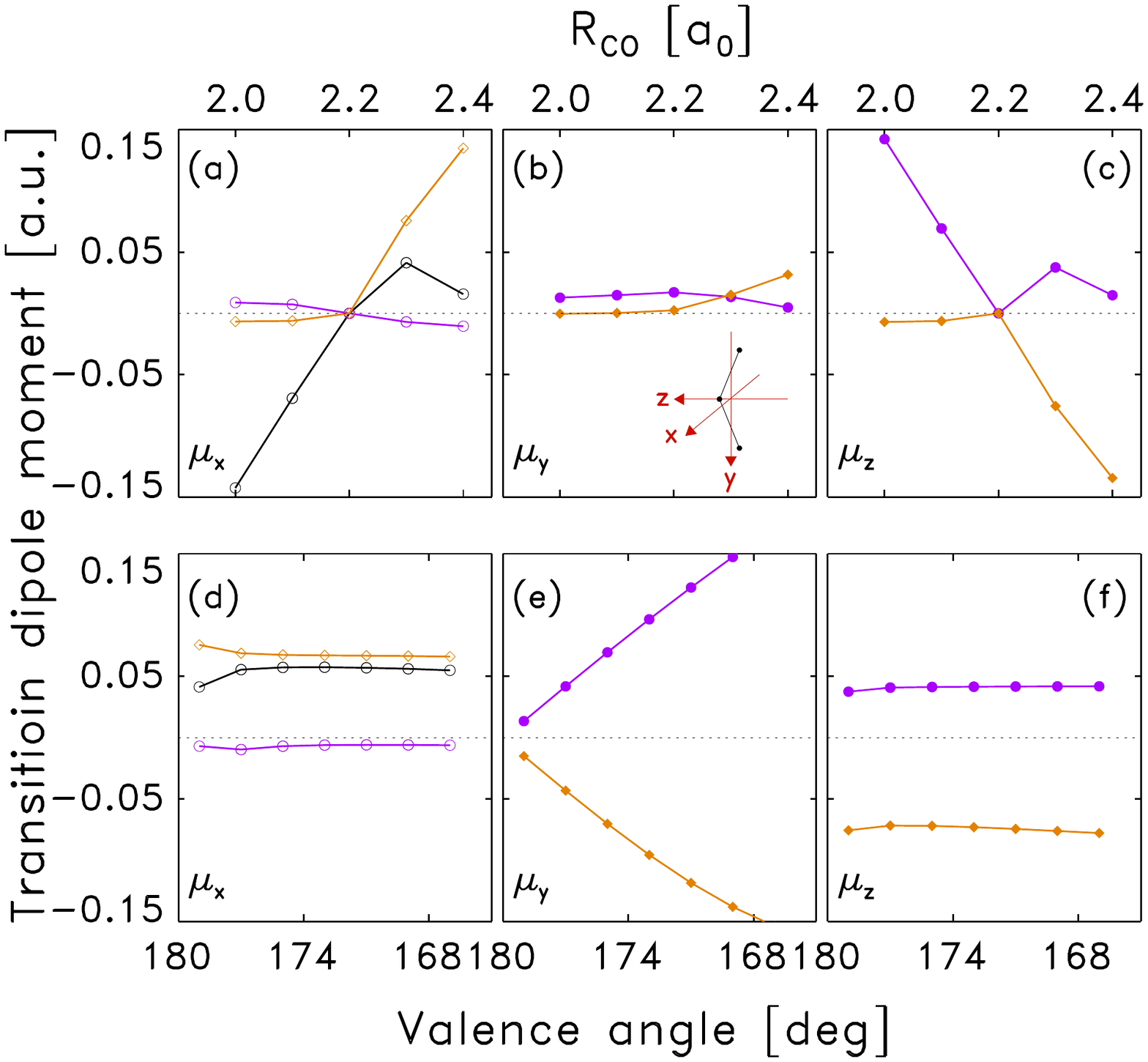}

\vspace{-1cm}

Fig.\ 3

\newpage
\mbox{ }
\vspace{1cm}

\includegraphics[angle=90,scale=0.6]{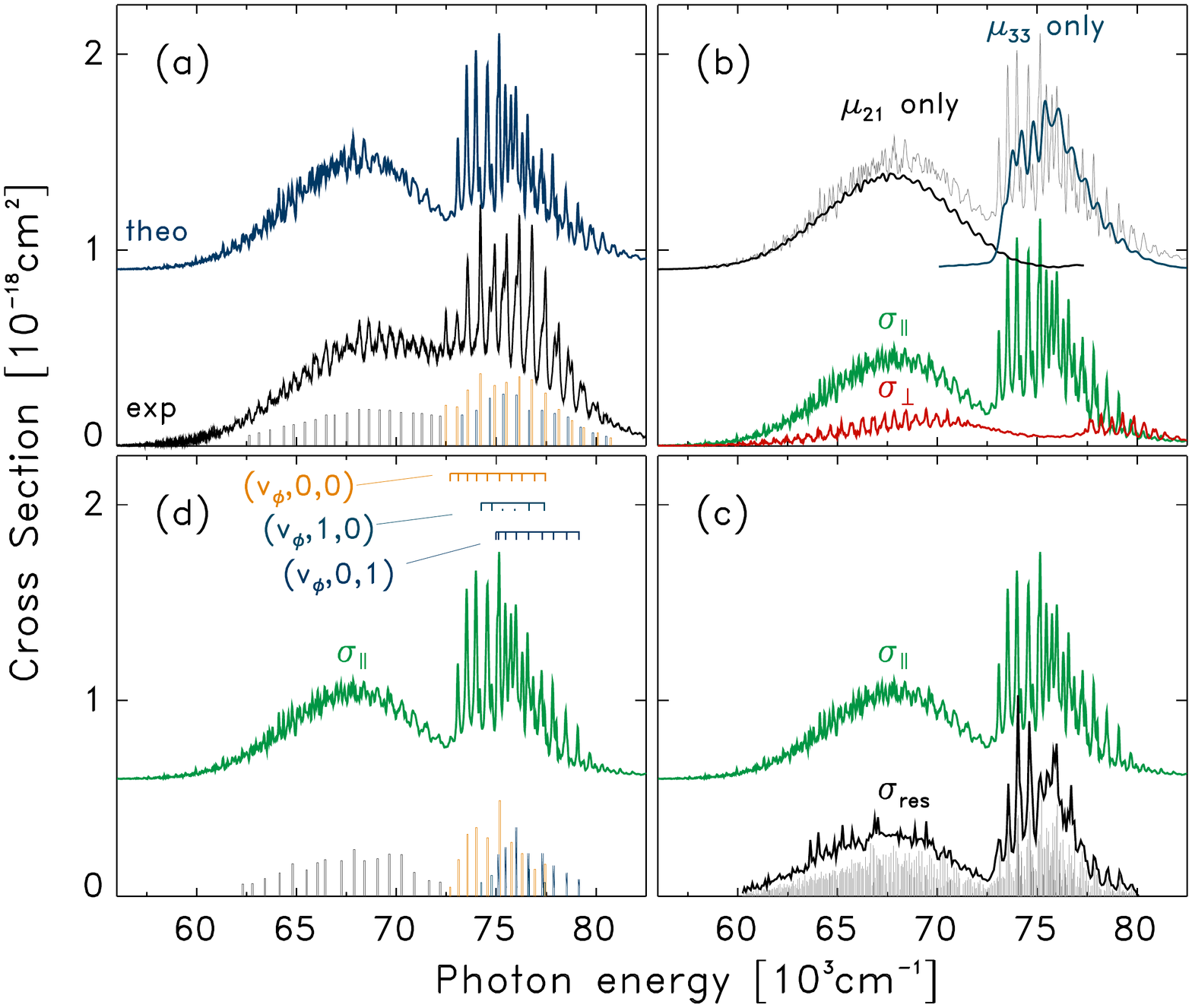}

\vspace{2cm}

Fig.\ 4

\newpage
\mbox{ }
\vspace{-1cm}

\includegraphics[angle=0,scale=0.7]{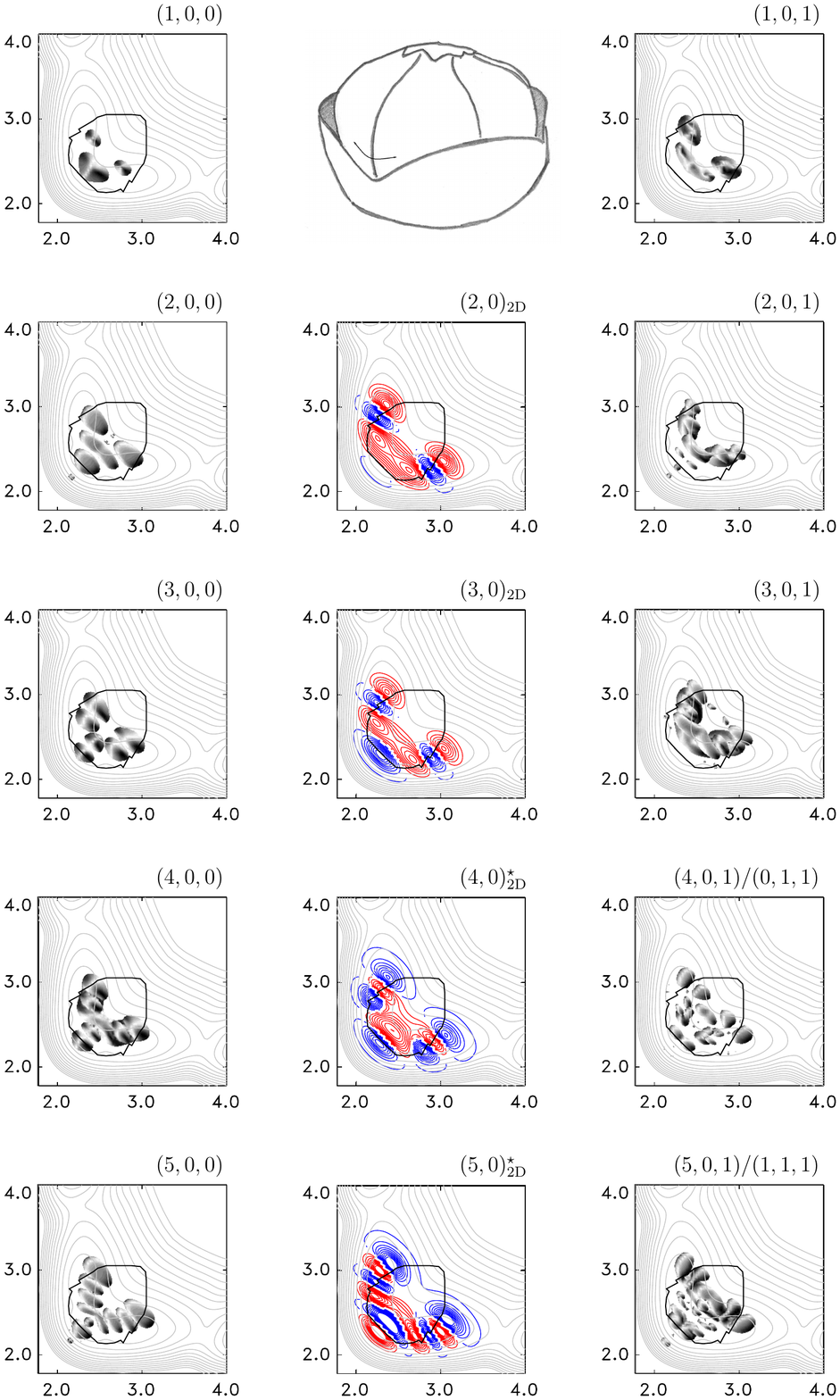}

\vspace{-1cm}

Fig.\ 5

\newpage
\mbox{ }
\vspace{1cm}

\includegraphics[angle=0,scale=0.7]{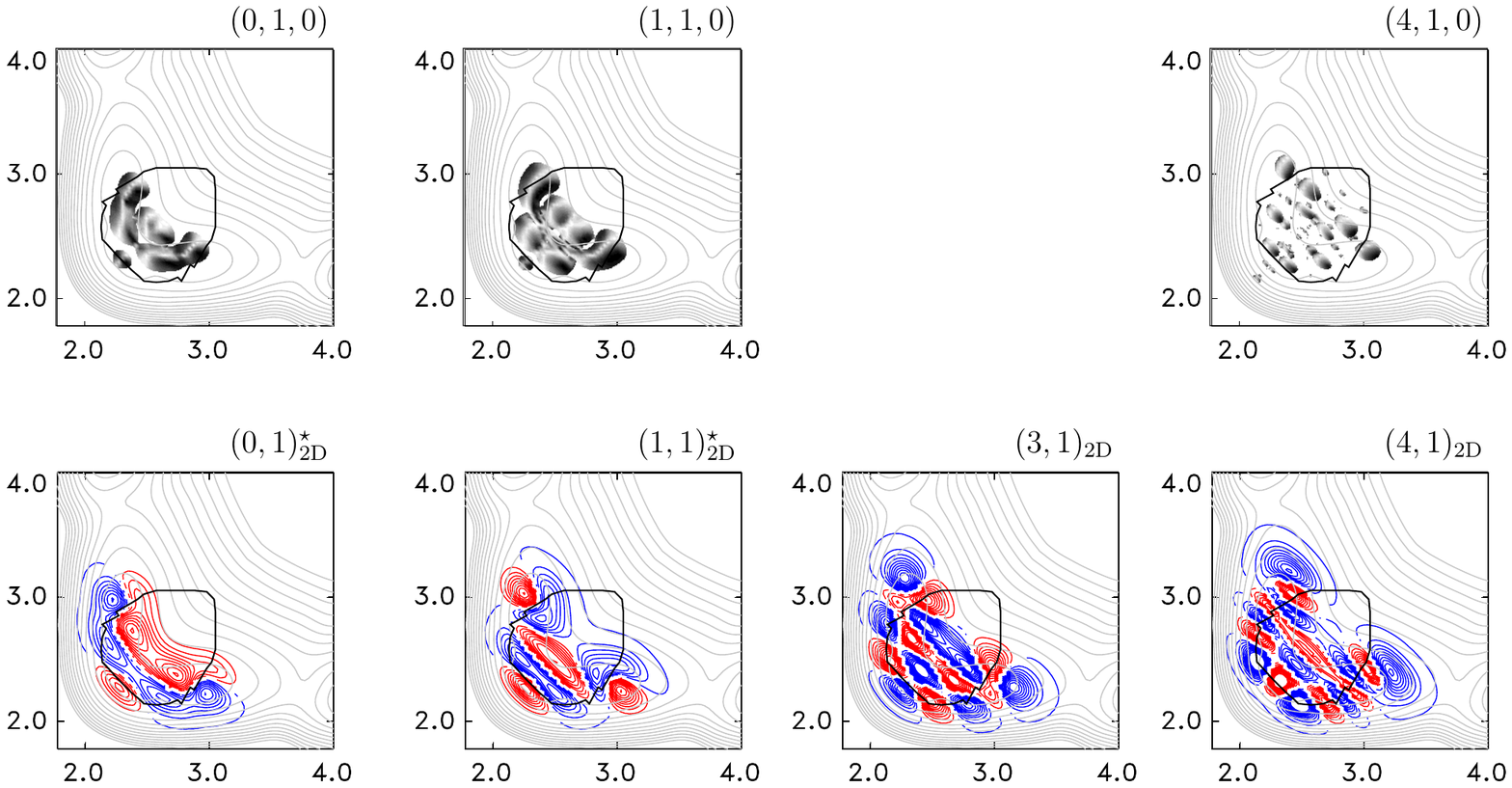}

\vspace{-2cm}

Fig.\ 6

\newpage
\mbox{ }
\vspace{1cm}

\includegraphics[angle=0,scale=0.7]{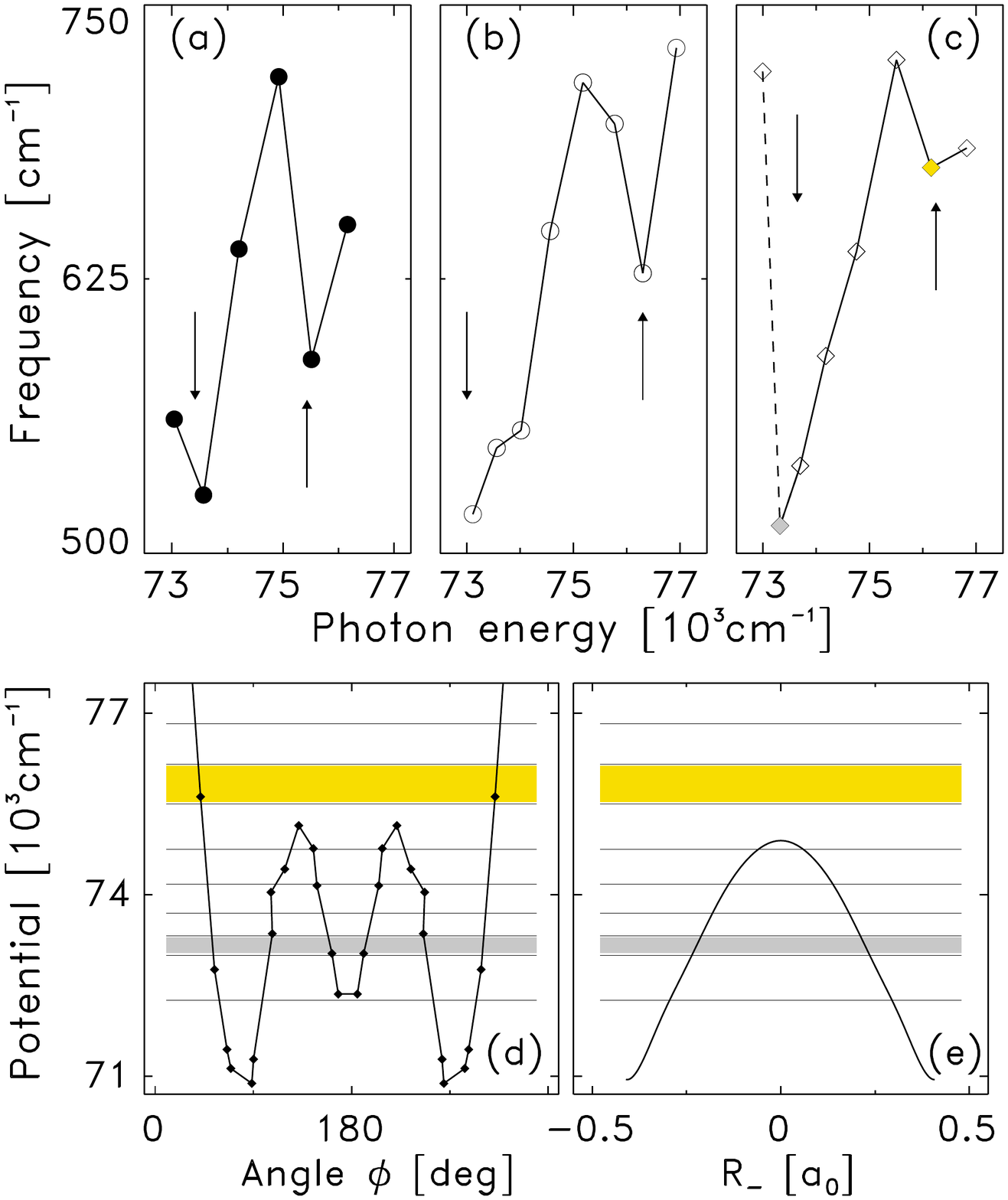}

\vspace{-1cm}

Fig.\ 7

\newpage
\mbox{ }
\vspace{1cm}

\includegraphics[angle=0,scale=0.7]{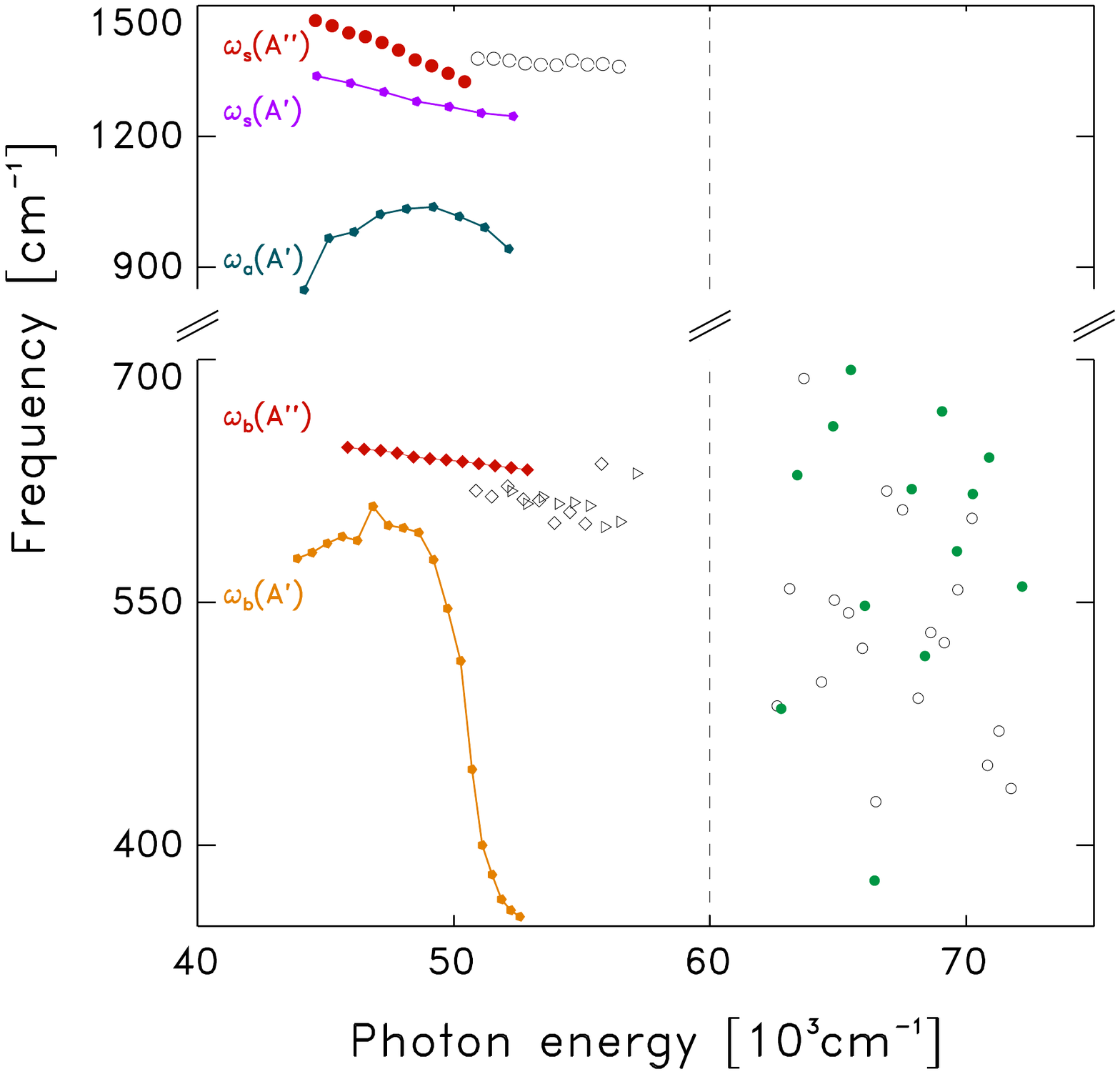}

\vspace{-1cm}

Fig.\ 8

\newpage
\mbox{ }
\vspace{1cm}

\includegraphics[angle=0,scale=0.7]{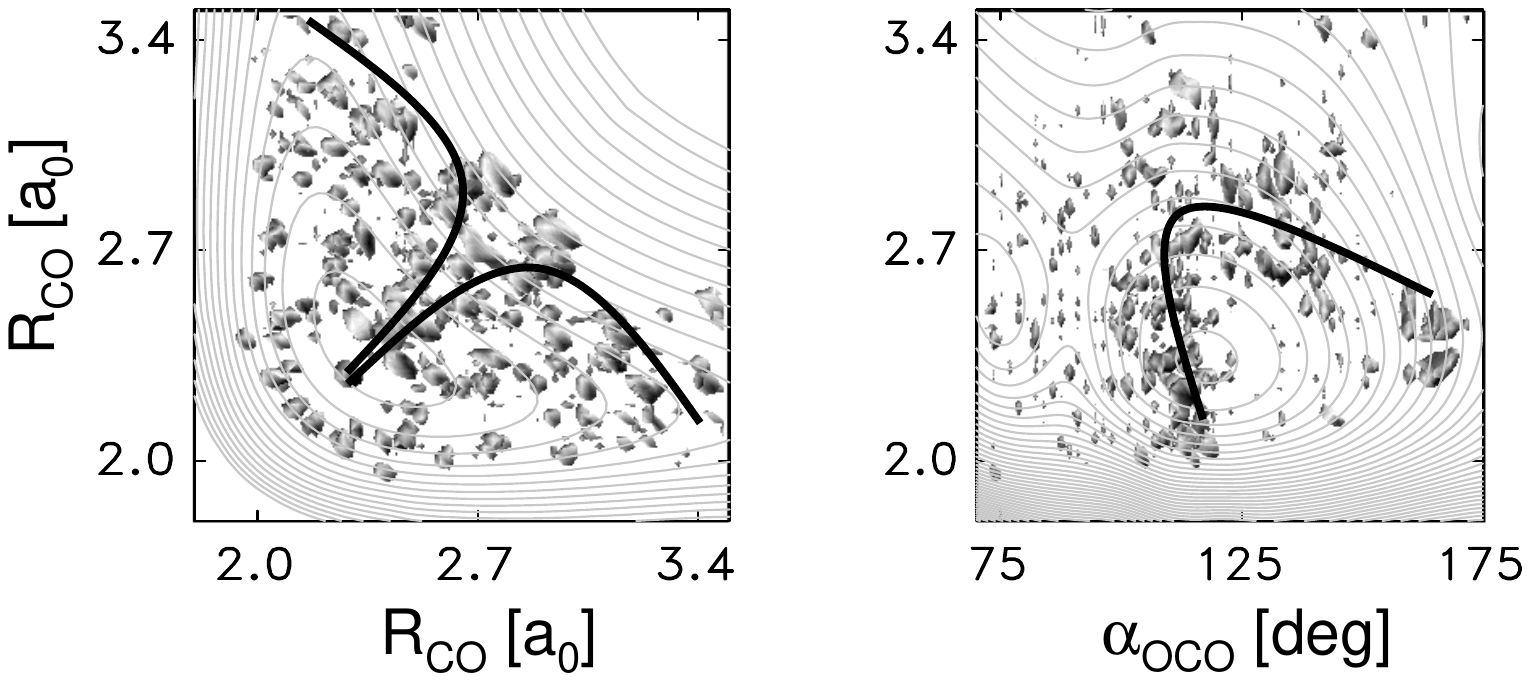}

\vspace{-1cm}

Fig.\ 9

\newpage
\mbox{ }
\vspace{1cm}

\includegraphics[angle=0,scale=0.7]{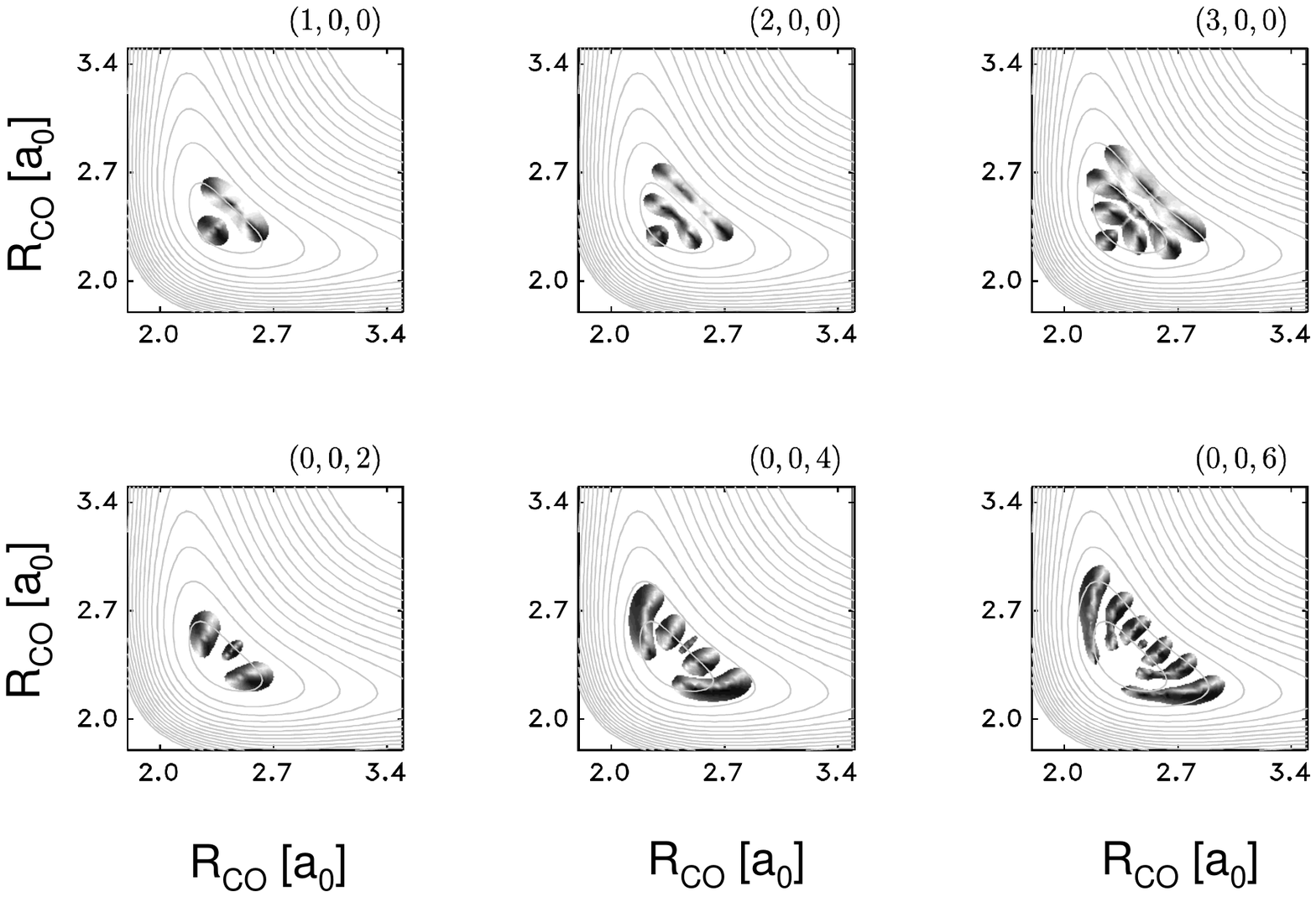}

\vspace{2cm}

Fig.\ 10

\newpage
\mbox{ }
\vspace{1cm}

\includegraphics[angle=0,scale=0.7]{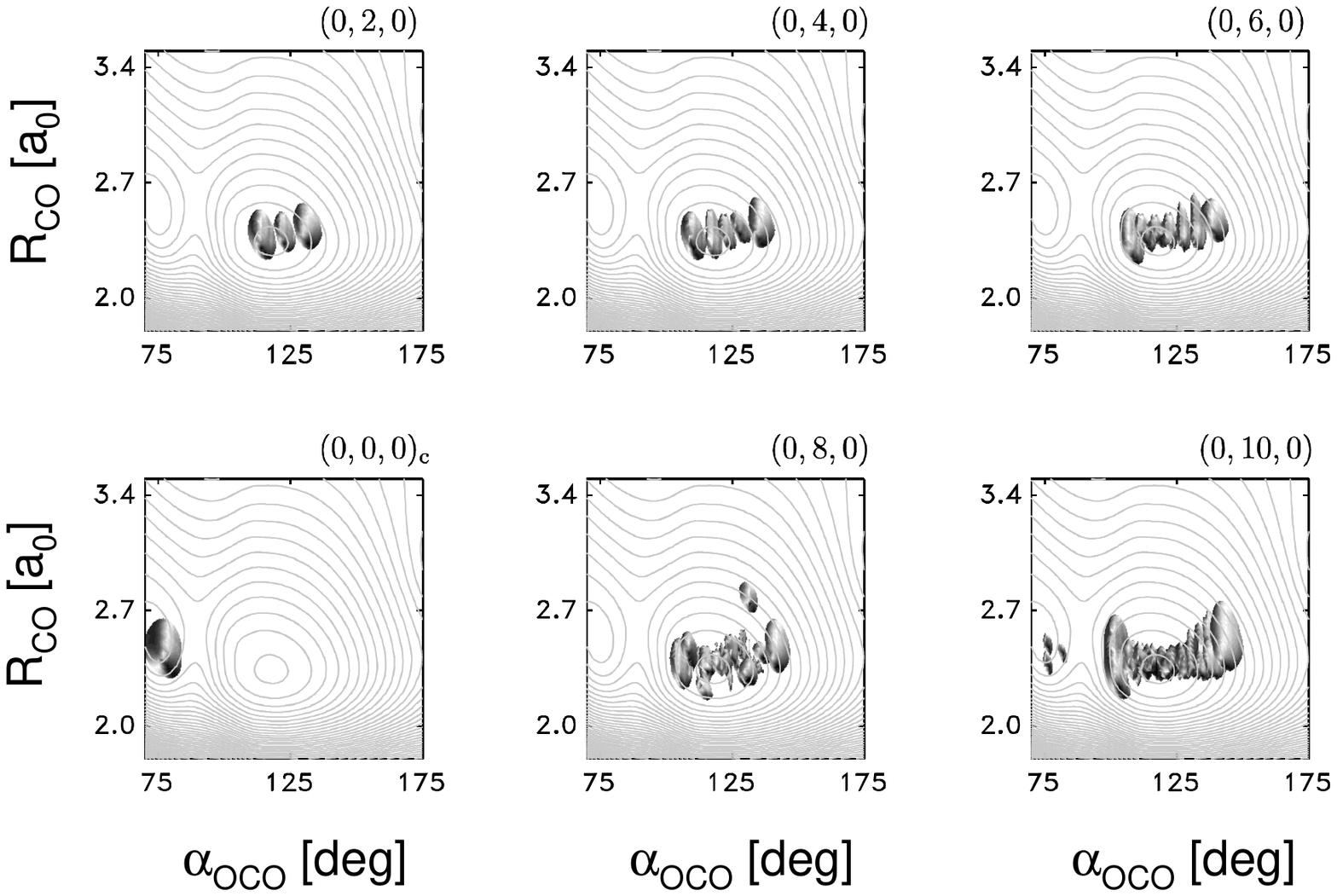}

\vspace{2cm}

Fig.\ 11

\newpage
\mbox{ }
\vspace{1cm}

\includegraphics[angle=0,scale=0.7]{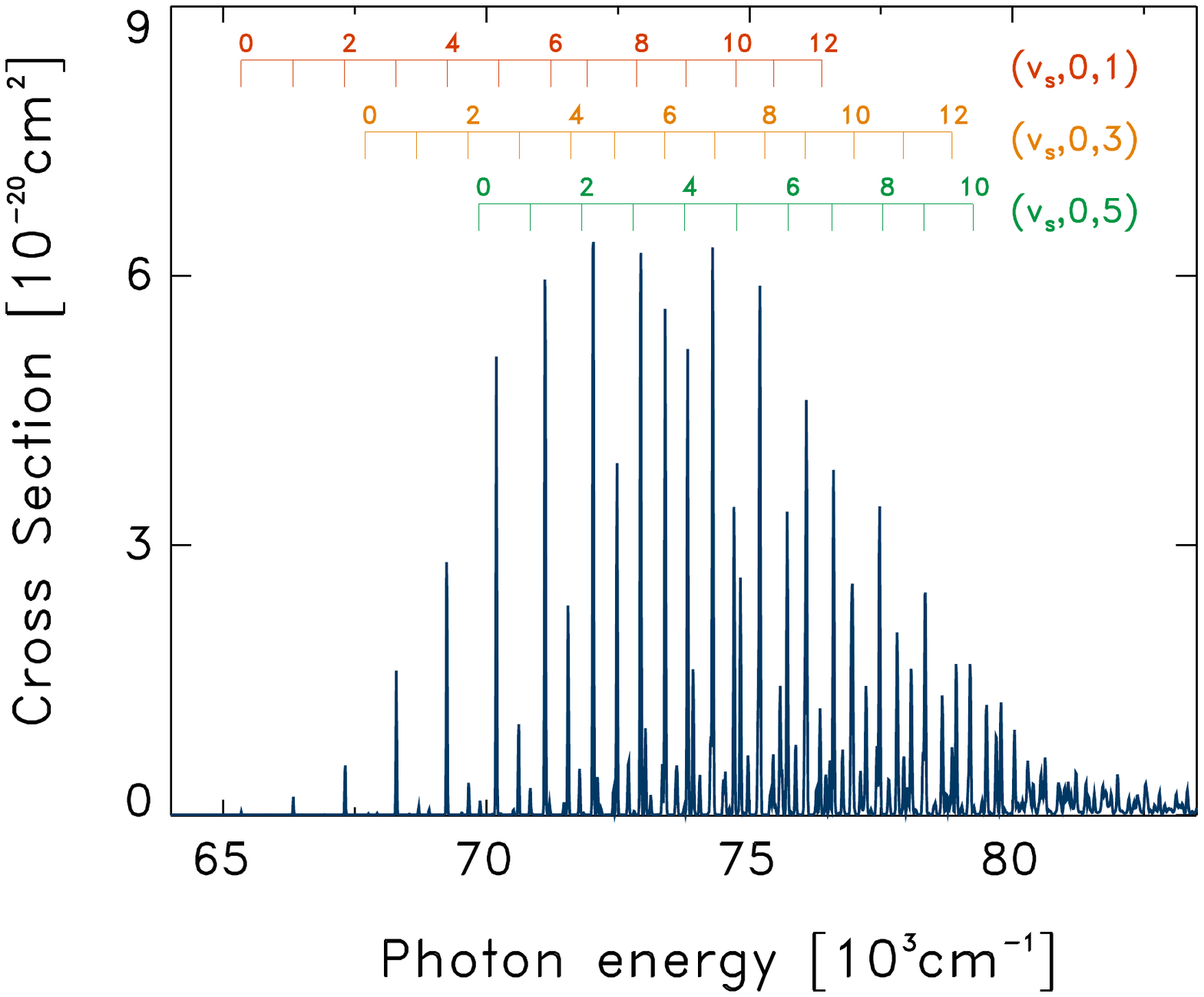}

\vspace{-1cm}

Fig.\ 12

\end{document}